\documentclass{aa}  
\usepackage{graphicx}
\usepackage{txfonts}

\begin{document}

   \title{Searching for $\delta$ Scuti-type pulsation and characterising northern pre-main-sequence field stars}
   \titlerunning{Searching for pre-main-sequence $\delta$ Scuti-type field stars}
   \authorrunning{Díaz-Fraile et al.}

   \author{D. Díaz-Fraile, E. Rodríguez \and P.J. Amado}

   \institute{Instituto de Astrofísica de Andalucía, CSIC, P.O. 3004, E-18080 Granada, Spain\\
              \email{dario@iaa.es}}

  \abstract
   {Pre-main-sequence stars (PMS) are objects evolving from the birthline to the zero-age main sequence (ZAMS). Given a mass range near the ZAMS, 
the temperatures and luminosities of pre-main-sequence and main-sequence stars are very similar. Moreover, their evolutionary tracks intersect one another causing some ambiguity in 
the determination of their evolutionary status. In this context, the detection and study of pulsations in pre-main-sequence stars is crucial for differentiating 
between both types of stars by obtaining information of their interiors via asteroseismic techniques.
  }
   {A photometric variability study of a sample of northern field stars, which previously classified as either PMS or Herbig Ae/Be objects, has been undertaken 
with the purpose of detecting $\delta$ Scuti-type pulsations. Determination of physical parameters for these stars has also been carried 
out to locate them on the HR diagram and check the instability strip for this type of pulsators.}
   {Multichannel photomultiplier and CCD time series photometry in the \textit{uvby} Str\"{o}mgren and \textit{BVI} Johnson bands 
were obtained during four consecutive years from 2007 to 2010. The light curves have been analysed, and a variability criterion has been 
established. Among the objects classified as variable stars, we have selected 
those which present periodicities above 4 d$^{-1}$, which was established as the lowest limit for $\delta$ Scuti-type pulsations in 
this investigation. Finally, these variable stars have been placed in a colour-magnitude diagram using the physical parameters 
derived with the collected \textit{uvby$\beta$} Str\"{o}mgren-Crawford photometry.}
   {Five PMS $\delta$ Scuti- and three probable $\beta$ Cephei-type stars have been detected. Two additional PMS $\delta$ Scuti stars are also confirmed in this work. 
Moreover, three new $\delta$ Scuti- and two $\gamma$ Doradus-type stars have been detected among the main-sequence objects used as comparison or check stars.}
   {}


   \maketitle
%
%________________________________________________________________

\section{Introduction}

Pre-main-sequence (PMS) stars are objects that evolve from the birthline, which is the locus in the HR diagram along which young stars first appear
as optically visible objects \citep{1983ApJ...274..822S}, to the zero-age main sequence (ZAMS). They are observed both in the field, mainly in the vicinity of molecular clouds, 
and in young open clusters with spectral types between late B and late M. In the case of PMS field stars, they are characterised by a high degree of activity, strong IR excess caused by 
the presence of circumstellar material, and emission lines in most cases. They show photometric and spectroscopic variability due to the photospheric activity 
and interaction with the circumstellar medium. These observational characteristics classify them as either T Tauri stars \citep{1945ApJ...102..168J,1962AdA&A...1...47H}  when their 
spectral types are later than F or as Herbig Ae/Be (HAEBE) objects \citep{1960ApJS....4..337H,1984A&AS...55..109F} when their spectral types are in the early F to late B range. \\

Close to the ZAMS, the evolutionary tracks of PMS and post-main\,sequence stars intersect each other several times \citep{1998A&A...332..958B}. Therefore, their evolutionary status 
is ambiguous at a given combination of effective temperature ($T_{\rm eff}$), luminosity ($L/L_{\odot}$), and mass ($M_{\odot}$). The PMS stars only differ from their counterparts on the 
main-sequence (MS) in their interiors, whereas their atmosphere properties are very similar \citep{1998ApJ...507L.141M}. \\

In their contraction phase towards the ZAMS, PMS stars with intermediate masses, i.e., between $\sim$1.5 and 4 M$_{\odot}$, cross the classical instability strip, where 
$\delta$ Scuti-type pulsations can be excited. Several pulsating PMS stars have been discovered and studied within the last few years 
\citep{2008ApJ...673.1088Z,2009A&A...502..239Z}. Their pulsation periods range between 18 minutes (80 d$^{-1}$; \citealt{2004MNRAS.352L..11A}) and about 6 hours (4 d$^{-1}$)
with amplitudes usually of only a few millimagnitudes \citep{2001A&A...366..178R}. As the pulsational properties depend on the inner structure of the stars, the detection of 
$\delta$ Scuti-type pulsations is particularly important to discriminate between both different evolutionary stages and the modelling of their interiors using asteroseismic techniques
 \citep{2007ApJ...671..581G,2009ApJ...704.1710G,2011ApJ...729...20Z,2013A&A...552A..68Z}. \\

The main aim of this work was to conduct a search for $\delta$ Scuti-type pulsations over a sample of northern field stars that are brighter than V $\approx$ 13.$^{m}$0, which are 
previously classified as either PMS or HAEBE stars. This previous classification comes from different sources; they mainly come from the \citet{1994A&AS..104..315T} and 
\citet{2003AJ....126.2971V} HAEBE catalogues and the list with all known PMS pulsators and pulsating candidate stars by \citet{2008ApJ...673.1088Z}. 
An initial sample of 63 stars was selected, which includes all PMS and HAEBE field stars known to date and visible from the
northern hemisphere. However, four stars (UX Ori, HD 34282, V436 Ori, and V351 Ori) were rejected from the sample because they had already been studied properly
in the literature. \\

The paper is organized as follows. The observations and data reduction are described in Sect. 2. Frequency analysis and results obtained are described in Sects. 3, 4 and 5. 
Sect. 6 is devoted to the determination of the stellar physical parameters and to locating the stars in a colour-magnitude diagram. The summary and the main conclusions are 
given in Sect. 7. 

\section{Observations and data reduction}

The sample was observed photometrically between the years 2007 and 2010 using the 1.5-m (T150) and 0.9-m (T90) telescopes at Sierra Nevada 
Observatory (SNO), Granada, Spain. In the case of T150, Johnson \textit{BVI} filters were used with CCD imaging for stars that are fainter than 
V $\approx$ 10.$^{m}$5. In the case of T90 for stars brighter than V $\approx$ 10.$^{m}$5, simultaneous multicolour Strömgren \textit{uvby} photometry was carried out using the six-channel 
\textit{uvby} photometer (Strömgren photometer) that is attached to this telescope \citep{1997A&A...324..959R}. 
This magnitude value was established as the limit for the T90 for providing sufficient precision in searching for very-low-amplitude pulsations.
Nevertheless, all stars (except RR Tau, see Table~\ref{tab:Dereddenedindices}) were observed with the Str\"{o}mgren photometer to obtain their \textit{uvby$\beta$} indices 
for calibration purposes. In Table~\ref{tab:ObjobI}, the list of ID numbers for each object is provided (ID numbers 7,8,11, and 29 correspond to UX Ori, HD 34282, V436 Ori, and V351 Ori, 
respectively, are not included in the Table, see Sect. 1) with \textit{V} magnitudes, coordinates (J2000), and spectral types obtained from the literature. Basic information 
about the observations carried out for the 59 PMS stars in our sample is also provided there. In total, photometric data have been
acquired on 136 nights (78 nights on T90 and 58 nights on T150) by obtaining 480 hours of observation (238 h on T90 and 197 h on T150) and collecting
12\,823 data points (4261 points on T90 and 8\,562 points on T150). Traditional criteria of proximity, similar brightness, and spectral types were applied 
when selecting the comparison and check stars in the case of observations with T90 and Str\"{o}mgren photometer. The same selection criteria were applied 
to T150 observations but for stars within the CCD FoV.\\

The atmospheric extinction correction was based on nightly coefficients determined from the main comparison star in each case. Magnitude differences 
were also calculated relative to the main comparison star by means of linear interpolation. For T90 data, instrumental magnitude differences 
were transformed into a standard \textit{uvby$\beta$} system following the procedure described in \citet{1997A&A...324..959R}. The T150 data were 
reduced with IRAF\footnote{The Image Reduction and Analysis Facility IRAF is
distributed by the National Optical Astronomy Observatories, which are
operated by Association of Universities for Research in Astronomy, Inc.,
under cooperative agreement with the National Science Foundation.} software package following the standard procedure. Dates of observations were converted to the Heliocentric Julian Date (HDJ),
and the zero point in time was set to the beginning of the observations for each observed star. \\

\begin{sidewaystable*}
\caption{Full sample of observed HAEBE and PMS field stars.}\label{tab:ObjobI}
\begin{minipage}[c]{0.5\textwidth}
\centering
\scalebox{0.60}{
\begin{tabular}{cccccccccc}
\hline\hline
ID &Name & $\alpha$ & $\delta$ & V & Spectral & Telescope & Season & Hours & Points \\
Number &      & (J2000)  & (J2000)  &  (mag)      &  Type    &SNO     &        & (nights)&\\
\hline \\

1      &VX Cas  & 00 31 30.68 & +61 58 50.97 & 11.30 & A0 &T150&Aug07& 3.10 (1)  & 220\\
       & ~      &             &              &       &    &T150&Nov07& 4.15 (1)  & 58  \\
2      &V594 Cas& 00 43 18.25 & +61 54 40.14 & 10.64 & Be &T150&Aug07& 6.03 (1)  & 392 \\
3      & PDS 004 & 03 39 00.56 & +29 41 45.70 & 10.74 & A1 &T150&Nov07& 2.21 (1)  & 68 \\
       & ~      &             &              &       &    &T150&Nov08& 3.96 (1)  & 208 \\
       & ~      &             &              &       &    &T90 &Dec09& 4.62 (1)  & 68  \\
4      & XY Per  & 03 49 36.32 & +38 58 55.60 &  9.44 &A2II&T90 &Oct07& 3.50 (1)  &  57 \\
       & ~      &             &              &       &    &T90 &Dec09& 2.38 (1)  &  37  \\
5      & AB Aur  & 04 55 45.84 & +30 33 04.29 &  7.06 &A0Vpe&T90&Oct07& 2.91 (1)  &  38 \\
6      & HD 31648& 04 58 46.26 & +29 50 36.98 &  7.73 &A3  &T90&Oct07& 3.02 (1)  & 39 \\
9      & HD 35187&  05 24 01.17 & +24 57 37.58 & 7.78 &A2e &T90&Oct07& 8.46 (2)  & 184 \\
       & ~      &              &              &      &    &T90&Dec07 & 1.41 (1)  & 37 \\
       & ~      &              &              &      &    &T90&Dec09 & 3.39 (1)   & 64  \\
10     &HD 287823& 05 24 08.05 & +02 27 46.89 & 9.67 & A0 &T90&Nov07& 2.91 (1)  & 32 \\
12     &HD 290409& 05 27 05.47 & +00 25 07.61 & 9.96 & B9 &T90&Nov07& 2.92 (1)  & 33 \\
13     &HD 290500& 05 29 48.03 & -00 23 43.16 & 11.04 & B8 &T150&Nov07& 2.33 (1) & 44 \\
       & ~     &                &              &       &    &T150&Nov08 & 3.49 (1)   &  119  \\
       & ~     &                &              &       &    &T150&Oct09 &  3.17 (1)  &  156  \\
14     &V1409 Ori& 05 30 19.03 & +11 20 19.90  & 9.30 & A1ab &T90&Oct07& 4.94 (1) & 69 \\
       & ~     &                &              &       &      &T90&Nov07 & 5.94 (2)  &  131  \\
       & ~     &                &              &       &      &T90&Dec09 &  4.07 (1)   &  68   \\
15     &HD 36112&  05 30 27.53 & +25 19 57.08 & 8.31  & A3e &T90&Oct07& 8.38 (2)  & 114 \\
       & ~     &                &              &       &     &T90&Dec07 & 1.32 (1)   &  18  \\
       & ~     &                &              &       &     &T90&Dec09 & 8.32 (1)  &  153   \\
16     &V1410 Ori& 05 31 57.24 & +11 17 41.46 & 9.43  & A3  &T90&Oct07& 4.94 (1)  & 63 \\
       &~     &                &              &       &     &T90&Nov07  & 2.72 (1)   &  28  \\
17     &HD 36408& 05 32 14.14 & +17 03 29.25 & 5.46   & B7IIIe&T90&Oct07&5.61 (1)  & 75  \\
       & ~     &               &              &        &       &T90&Nov07&3.11 (1)  &  75  \\
18     &V1271 Ori& 05 35 09.60 & +10 01 51.51 & 10.00 & A5    &T90&Oct07& 5.62 (1) & 73  \\
19     &V380 Ori& 05 36 25.43 & -06 42 57.69 &  10.70 & A1e   &T150&Nov07& 2.81 (1) & 45  \\
20     &HD 290770& 05 37 02.45 & -01 37 21.36 & 9.30  & B8    &T90&Nov07& 2.42 (1) & 31  \\
21     &BF Ori& 05 37 13.26 & -06 35 00.58 & 10.30 & A5II-IIIe&T90&Oct07& 1.58 (1)  &  13 \\
       &~     &             &              &       &           &T90&Nov07&9.61  (2) &  138  \\
22     &HD 37357& 05 37 47.08 &-06 42 30.25 & 8.85 & A0e       &T90&Oct07&5.58 (1)  &  58 \\
23     &V1247 Ori& 05 38 05.25 & -01 15 21.67 & 9.82 & A5III &T90&Nov07&2.33 (1)  & 30  \\
24     &V1788 Ori& 05 38 14.50 & -05 25 13.30 & 9.86 & B9Ve  &T90&Nov07&5.13 (1)  & 57  \\
25     &RR Tau& 05 39 30.52 & +26 22 26.97 & 10.90 & A2II-IIIe&T150&Nov07& 1.26 (1) &  34 \\
       &~     &             &              &       &          &T150&Nov08        &1.86 (1)  &  70  \\
26     &V350 Ori& 05 40 11.77 & -09 42 11.05 & 10.40 & A0     &T150&Jan08& 5.24 (1) &  270 \\
       &~     &               &              &       &        &T150&Nov08        &2.23 (1)  & 83  \\
27     &HD 37806& 05 41 02.29 & -02 43 00.73 & 7.93  & A0     &T90&Dec07& 5.30 (1) & 67  \\
       & ~     &               &              &       &        &T90&Feb10        &1.71 (1)  & 21   \\
28     &HD 38120& 05 43 11.89 & -04 59 49.90 & 9.08  & A0     &T90&Dec07& 5.27 (1)  & 66 \\
       &~     &               &              &       &        &T90&Feb10 & 1.63 (1) &  20   \\
30     &HD 249879& 05 58 55.78 & +16 39 57.37 & 10.64 & B8    &T150&Jan08& 8.04 (1) & 649  \\
       &~     &                &              &       &       &T150&Nov08& 3.69 (1)  & 295    \\
       &~     &                &              &       &       &T150&Apr10& 0.86 (1)  & 74    \\
31     &V791 Mon& 06 02 14.88 &-10 00 59.50 & 9.60 & Be       &T90&Feb08& 5.38 (2)  &  149 \\
       &~     &               &             &      &          &T90&Feb10& 2.03 (1) & 31  \\
32     &HD 250550& 06 01 59.00 & +16 30 56.73 & 9.57 & B7e    &T90&Feb08& 4.86 (2) & 118 \\
       &~     &               &              &       &        &T90&Feb10& 2.24 (1)  & 35 \\
33     &AE Lep& 06 03 37.06  & -14 53 02.50 & 11.00 & Ae      &T150&Jan08& 5.27 (2)  &  358 \\
       &~     &              &              &       &         &T150&Nov08        &4.05 (1)  & 185   \\
34     &PDS 126& 06 13 36.20 & -06 25 01.00 & 11.82 & A7V     &T150&Jan08& 6.61 (1) &  218 \\
35     &HD 50083& 06 51 45.75 & +05 05 03.86 & 6.91 & B2Ve    &T90&Dec07& 5.36 (2) & 115  \\
       &~     &               &              &      &         &T90&Feb08        &2.84 (1)   & 101  \\
       &~     &               &              &      &         &T90&Feb10        &3.08 (1)  &  45 \\
36     &HD 52721& 07 01 49.51 & -11 18 03.32 & 6.58 & B2Ve?   &T90&Feb08& 8.40 (2) & 112 \\
       &~     &               &              &      &         &T90&Feb10        & 1.00 (1) & 9    \\
\\
\hline
\end{tabular}
}
\end{minipage}
\begin{minipage}[c]{0.5\textwidth}
\centering
\scalebox{0.60}{
\begin{tabular}{cccccccccc}
\hline\hline
ID &Name & $\alpha$ & $\delta$ & V & Spectral & Telescope & Season & Hours & Points \\
Number &      & (J2000)  & (J2000)  &  (mag)      &  Type    &SNO     &        & (nights)&\\
\hline \\
37     &HT Cma& 07 02 42.53 & -11 26 11.81 & 11.87 & A0       &T150&Feb11& 3.42 (1) &  149 \\
38     &HU Cma& 07 04 06.70 &-11 26 08.61 & 12.09 & B9e       &T150&Feb08& 4.06 (1) & 187  \\
39     &HD 53367& 07 04 25.53 & -10 27 15.74 & 7.04 & B0IV:e  &T90&Feb08& 8.41 (2)  & 117  \\
       &~     &               &              &      &         &T90&Feb10        & 1.00 (1) & 9   \\
40     &PDS 241& 07 08 38.80 & -04 19 08.00 & 12.06 & B0?     &T150&Jan08&  6.03 (2) & 151  \\
       &~     &              &              &       &         &T150&Nov08& 5.52 (1)  & 196    \\
       &~     &              &              &       &         &T150&Apr10& 1.21 (1)   & 60   \\
41     &HD 141569& 15 49 57.75 & -03 55 16.34 & 7.13 & B9.5e  &T90&Apr07& 5.54 (1)  &  93 \\
       &~     &                &             &       &        &T90&Feb10        & 1.55 (1) &  27   \\
42     &VV Ser& 18 28 47.86 & +00 08 39.76 & 11.60 & B6       &T150&Apr07& 7.01 (3)  & 136  \\
       &~     &             &              &       &          &T150&Apr10 & 1.75 (1)   &69    \\
43     &V431 Sct& 18 29 25.69 & -06 04 37.29 & 10.50 & Bpe    &T90&Jul07& 6.06 (2) & 85  \\
       &~     &             &                &       &        &T150&Aug07       &4.92 (2)  &204    \\
       &~     &             &                &       &        &T150&Apr10        &1.45 (2)  &41   \\
44     &HD 174571& 18 50 47.18 & +08 42 10.09 & 8.84 & B3V:pe &T90&May07& 4.15 (1) & 32  \\
       &~     &               &               &      &        &T90&Jul07        & 3.92 (1) & 60    \\
       &~     &              &                &      &        &T90&Aug09        & 3.43 (1)  & 59   \\
       &~     &              &               &       &        &T90&Apr10        & 2.62 (1) & 25   \\
45     &HD 179218& 19 11 11.25 & +15 47 15.64 & 8.84 & B9e    &T90&May07& 6.13 (2) & 47  \\
       &~     &               &               &      &        &T90&Jul07       & 3.92 (1) &  60   \\
       &~     &               &               &      &        &T90&Apr10        & 2.59 (1)  &  24   \\
46     &WW Vul& 19 25 58.75 & +21 12 31.28 & 10.51 & A3e      &T150&May07& 2.73 (1)  & 129  \\
       &~     &             &              &       &          &T150&Aug07       &  6.61 (1) & 352    \\
       &~     &             &              &       &          &T150&Aug09        & 1.15 (1)  & 63   \\
       &~     &             &              &       &          &T150&Apr10        & 1.73 (1) &  84   \\
47     &PX Vul& 19 26 40.26 & +23 53 50.85 & 11.67 & F0V:e    &T150&Aug07& 4.52 (1) & 127  \\
       &~     &             &              &       &          &T150&Aug09        & 2.78 (2) & 80   \\
       &~     &             &              &       &          &T150&Apr10        & 1.72 (1)  & 61   \\
48     &PDS 581& 19 36 18.91 & +29 32 50.00 & 11.67 & B0.sIV  &T150&Aug07& 4.55 (1) & 80  \\
       &~     &             &               &       &         &T150&Sep07        & 5.34 (1)  & 147    \\
       &~     &             &              &        &         &T150&Apr10        & 2.05 (1)  & 75    \\
49     &HD 190073& 20 03 02.51 & +05 44 16.67 & 7.82 & A2IVpe &T90&Jul07& 2.97 (1) & 45 \\
50     &V1685 Cyg& 20 20 28.24 & +41 21 51.56 & 10.70 & B2    &T150&Aug07& 4.83 (1) & 225  \\
       &~     &                &            &        &        &T150&Aug09        & 5.24 (2)  & 238    \\
51     &HD 200775& 21 01 36.92 & +68 09 47.76 & 7.42 & B2Ve   &T90&May07& 4.39 (1) & 49  \\
       &~     &               &             &        &        &T90&Jul07      &  4.20 (1) & 46   \\
52     &HD 203024& 	21 16 03.02 & +68 54 52.13& 8.88 & A    &T90&May07& 4.39 (1) & 50  \\
       &~     &               &              &         &      &T90&Jul07        &  4.20 (1)  & 47   \\
53     &BD +65.1637& 21 42 50.18 & +66 06 35.12 & 10.17 & B2nne &T90&Jul07& 8.91 (2) & 146  \\
54     &V1578 Cyg& 21 52 34.10 & +47 13 43.61 & 10.16 & B9.5Ve &T90&Jul07& 9.87 (2) & 165  \\
       &~     &               &              &        &        &T150&Aug07        &7.22  (1) & 606   \\
55     &BH Cep& 22 01 42.87 & +69 44 36.53 & 10.8 & F5IV       &T150&Aug07& 4.99 (1) & 261  \\
       &~     &             &              &      &            &T150&Nov07        & 3.77 (1)   & 67    \\
       &~     &             &               &     &            &T150&Aug09        & 2.57  (1) & 106    \\
56     &SV Cep& 22 21 33.20 & +73 40 27.07 & 10.10 & Ae        &T90&Jul07& 4.29 (1) &  72 \\
57     &V1080 Tau& 04 40 32.64 & +24 26 31.30 & 10.50 & G0      &T150&Nov07& 5.17 (1) & 98 \\
58     &CO Ori& 05 27 38.34 & +11 25 38.97 & 10.60 & G5Vpe     &T150&Nov07& 5.01 (1) & 94  \\
       &~     &             &              &       &           &T150&Nov08        & 3.14 (1)  & 115    \\
59     &V1650 Ori& 05 29 11.44 & -06 08 05.41 & 10.43 & F7     &T90&Nov07& 5.13 (1) & 57  \\
60     &RY Ori& 05 32 09.94 & -02 49 46.79 & 10.80 & F7        &T150&Nov07& 2.91 (1) & 36  \\
61     &HD 36910& 05 35 58.47 & +24 44 54.10 & 10.70 & F2IVe   &T90&Oct07& 3.63 (1) & 58  \\
62     &HD 53240& 07 03 57.45 & -10 07 25.55 & 6.43 & B9IIIn   &T90&Feb08& 8.28 (2) & 106  \\
63     &HD 261387& 06 39 20.79 & +09 33 51.00& 10.60 & A2V     &T90&Dec07& 6.87 (1) & 110  \\
       &~     &              &              &        &         &T150&Jan08        &6.14 (2)  &429    \\
       &~     &              &              &        &         &T150&Nov08        &2.98 (1)  &169    \\
       &~     &              &              &        &         &T150&Apr10        &2.40 (1) &181    \\
\\
\\
\\
\\
\hline
\end{tabular}
}
\end{minipage}
\end{sidewaystable*}

\section{Frequency analysis}
The HAEBE stars usually show large variations in luminosity from one year to another with an irregular (or at least not periodic) pattern due to the presence of circumstellar material 
\citep{1971AN....292..221W,1994A&A...292..165G,1999AJ....118.1043H,2001A&A...379..564O}. This long-term variability was removed from the whole data set for each star fitting a 
low-order Legendre polynomial. The residuals were then used to study their pulsational behaviour.\\

The program package PERIOD04 \citep{2005CoAst.146...53L} was used to perform the frequency analysis. The time interval between samples is about 3 min (T90) or 5 min (T150),
which leads to a Nyquist frequency of $f_{\rm Nyq}\sim$250 d$^{-1}$ or $f_{\rm Nyq}\sim$150 d$^{-1}$, respectively. The frequency spectra were first studied in the 
\textit{v} and \textit{B} bands because of the higher flux and pulsational amplitudes in A-F stars. The frequency range was selected from 0 to 100 d$^{-1}$ 
to include the highest frequencies predicted by models and observed in a PMS star \citep{2013MNRAS.428.2596C}. The 1$\sigma$ noise level was determined by computing the mean noise amplitude
in that range. Based on this value, 3$\sigma$ and 4$\sigma$ levels were calculated in each case. 
Our variability criterion was established as follows: \\

-Peaks with an amplitude larger than 4$\sigma$ level are considered as real peaks, and these stars as variable stars.

-Peaks between 3$\sigma$ and 4$\sigma$ levels are considered as doubtful, and the variability of the star as uncertain.

-Peaks below 3$\sigma$ level are considered as not significant and the star as not variable.\\
 
In the particular cases of stars with their main frequency peaks showing amplitudes between 3$\sigma$ and 4$\sigma$ levels, the \textit{Time Resolved
Image Photometry Package} (TRIPP, \citealt{2003BaltA..12..167S}) was also used to check the results obtained with PERIOD04. This package computes a Lomb-Scargle periodogram \citep{1982ApJ...263..835S}
instead of a Fourier amplitude spectrum. The program TRIPP provides a statistical estimate of the probability of a peak being real by using confidence levels for the relative power.
It was established that those peaks between 3$\sigma$ and 4$\sigma$ levels with a confidence $\geq$ 95$\%$ are considered as real 
peaks, and the star would be considered as variable. Following this criterion, we have found that 26 stars from our sample show some type of variability.
We note that our limit of 4$\sigma$ is similar to that with an amplitude signal-to-noise ratio S/N = 4.0, which is commonly accepted for this type of works \citep{1993A&A...271..482B,1999A&A...349..225B}. \\

Tables~\ref{tab:StarsI} and ~\ref{tab:StarsII} show the results obtained from the frequency analysis: object name, spectral type, visual magnitude, frequencies detected, amplitudes, 
mean noise level, and signal-to-noise ratio. Table~\ref{tab:StarsII} includes the results obtained with \textit{TRIPP} that shows the probability that the frequency detected would be 
real. In all cases, the frequencies, amplitudes, and S/N values listed correspond to the analysis obtained from Johnson \textit{B} (T150) or Strömgren \textit{v} (T90) bands. 
Frequency and amplitude errors are indicated in the parentheses that correspond to the last digits. They are the formal error bars given by PERIOD04, which are computed using the 
formulae derived by \cite{1999DSSN...13...28M}. Nevertheless, the calculated errors in frequency are very small because the stars were observed during very few 
nights spanning in time over different years in most of the stars in our study. This leads to unreliably small error bars in frequency. In these cases, we assume that the errors in our frequency determinations are 
much larger, typically of about 0.01 d$^{-1}$. \\ 

To establish the type of variability and discriminate between an intrinsic pulsation of the star and an external effect, 
such as luminosity variations produced by the circumstellar material commonly present around this type of HAEBE star, a more detailed analysis of all the different bands 
(\textit{uvby} or \textit{BVI}) was carried out and is described in the next section. 

\begin{table}
\caption{Stars of the sample with S/N$\geq$4.0.} 
\centering
\scalebox{0.70}{
\begin{tabular}{cccccccc}
\hline\hline
 ID &Name & Spectral &  V& Frequency & Amplitude  &     N     &  S/N  \\      
 number &     &  Type    & (mag)   &  (d$^{-1}$)    &  (mmag)    &   (mmag)  &       \\      
\hline
\\

2         &V594 Cas  & Be & 10.64 & 3.98 (11) & 9.7 (5) & 1.2 & 8.1  \\
3         &PDS 004   & A1 & 10.74 & 53.27 (1) & 8.5 (7) & 2.0 & 4.2  \\
4         &XY Per    & A2IIv & 9.44 & 5.27 (1) & 4.4 (5) & 0.8 & 5.5   \\
9         &HD 35187   & A2e & 7.78 & 63.87 (1) & 3.3 (4) & 0.7 & 4.7  \\
14        &V1409 Ori & AIab & 10.20 & 45.35 (1) & 7.8 (7) & 1.2 & 6.5  \\
          &~         &      &       &  2.43 (1) & 6.5 (5) & 1.2 & 5.4   \\
15        &HD 36112   & A3e  &  8.31 & 28.36 (1) & 3.7 (3) & 0.5 & 7.4  \\
          &~         &      &       & 33.00 (1) & 2.6 (3) & 0.5 & 5.2   \\
32        &HD 250550  & B7e  & 9.57  &  5.48 (1) & 4.3 (4) & 0.8 & 5.4   \\
36        &HD 52721   & B2Ve? & 6.58 &  3.51 (1) & 80.7 (23) & 9.0 & 9.0  \\
40        &PDS 241   & B0? & 12.06  & 2.86 (1) & 4.6 (4) & 0.8 & 5.7  \\
42        &VV Ser    & A2e & 11.60 & 9.65 (4) & 2.8 (4) & 0.5 & 5.6  \\
43        &V431 Sct  & Bpe & 10.50 & 11.66 (5) & 6.8 (9) & 1.0 & 6.8  \\
44        &HD 174571  & B3V:pe & 8.84 & 4.51 (1) & 4.7 (4) & 0.6 & 7.8  \\
46        &WW Vul    & A3e & 10.51 & 4.30 (1) & 4.8 (5) & 0.9 & 5.3  \\
48        &PDS 581   & B0.sIV & 11.67 & 6.83 (1) & 4.8 (5) & 0.5 & 9.6  \\
50        &V1685 Cyg & B3     & 10.70 & 7.10 (1) & 4.0 (4) & 0.8 & 5.0   \\
55        &BH Cep    & F5IV   & 10.80 & 5.57 (1) & 2.9 (3) & 0.5 & 5.8   \\
57        &V1080 Tau & G0     & 10.50 & 2.86 (7) & 22.5 (6) & 3.0 & 7.5  \\
59        &V1650 Ori & F7     & 10.43 & 3.04 (12) & 67.0 (30) & 9.0 & 7.4   \\
61        &HD 36910   & F2IVe  & 10.70 & 2.81 (4) & 68.0 (40) & 8.1 & 8.4   \\
63        &HD 261387  & A2V    & 10.60 & 34.67 (1) & 2.7 (3) & 0.6 & 4.5  \\
\\
\hline
\\
\end{tabular}
}
\label{tab:StarsI}
\end{table}

\begin{table}
\caption{Stars of the sample with 3.0$<$S/N$<$4.0.} 
\centering
\scalebox{0.62}{
\begin{tabular}{ccccccccc}
\hline\hline
 ID &Name & Spectral & V & Frequency & Amplitude  &     N     &  S/N  & Prob.   \\  
 number &     &  Type    & (mag)   &  (d$^{-1}$)    &  (mmag)    &   (mmag)  &       & (\%)    \\ 
\hline
\\

6   &HD 31648   & A3 & 7.73 & 5.60 (70) & 3.8 (6) & 1.2 & 3.2 & 98  \\
13  &HD 290500  & B8 & 11.04 & 8.18 (1) & 4.8 (7) & 1.5 & 3.2 & > 99 \\
17  &HD 36408   & B7IIIe & 5.46 & 15.47 (2) & 2.0 (4) & 0.6 & 3.3 & 20   \\
21  &BF Ori    & A5II-IIIeV & 10.30 & 5.62 (1) & 11.4 (16) & 3.3 & 3.4 & > 99  \\
26  &V350 Ori  & A0 & 10.40 & 57.08 (1) & 3.3 (4) & 0.9 & 3.7 & > 99   \\
35  &HD 50083   & B2III & 6.91 & 6.45 (1) & 2.7 (3) & 0.7 & 3.8 & > 99    \\
45  &HD 179218  & B9e  &  8.84 & 4.81 (1) & 1.5 (3) & 0.4 & 3.7 & 0  \\
58  &CO Ori    & G5Vpe & 10.60 & 5.34 (1) & 2.7 (4) & 0.7 & 3.8 & > 99    \\
\\
\hline
\\
\end{tabular}
}
\label{tab:StarsII}
\end{table}

\section{Pulsating PMS field stars}

\begin{figure*}
   \centering
   \includegraphics[width=9cm]{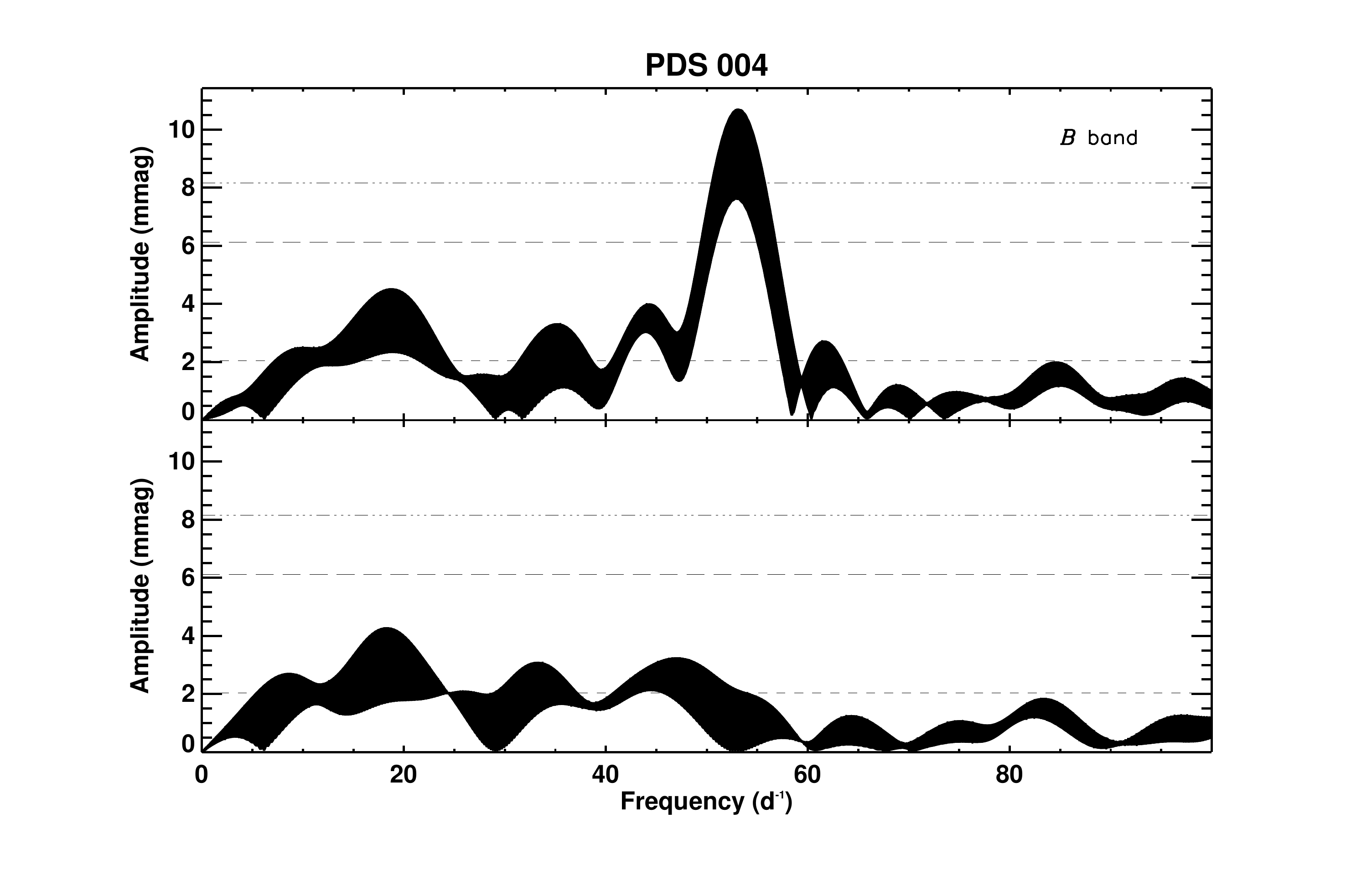}
  \includegraphics[width=9cm]{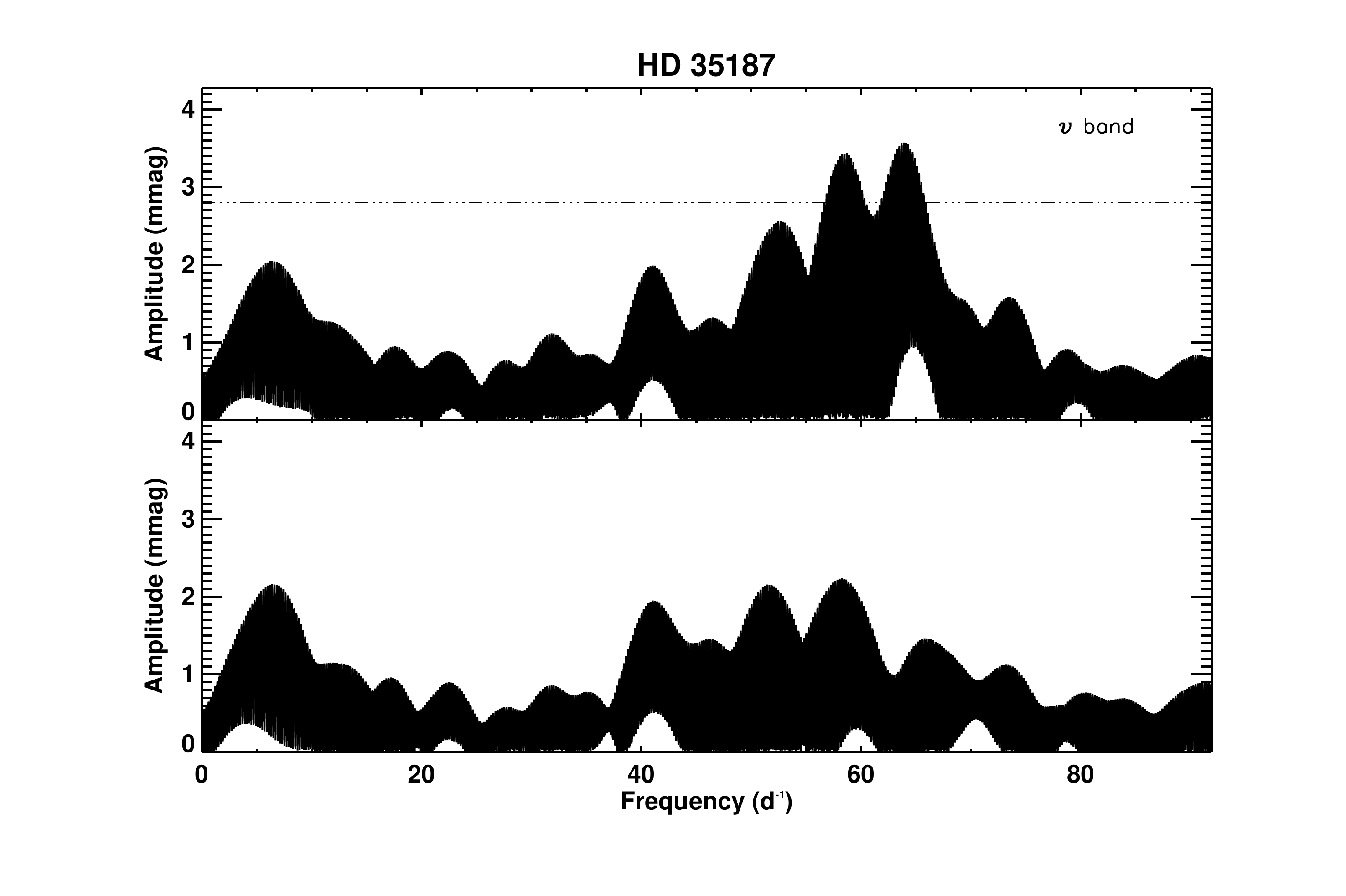}
   \includegraphics[width=9cm]{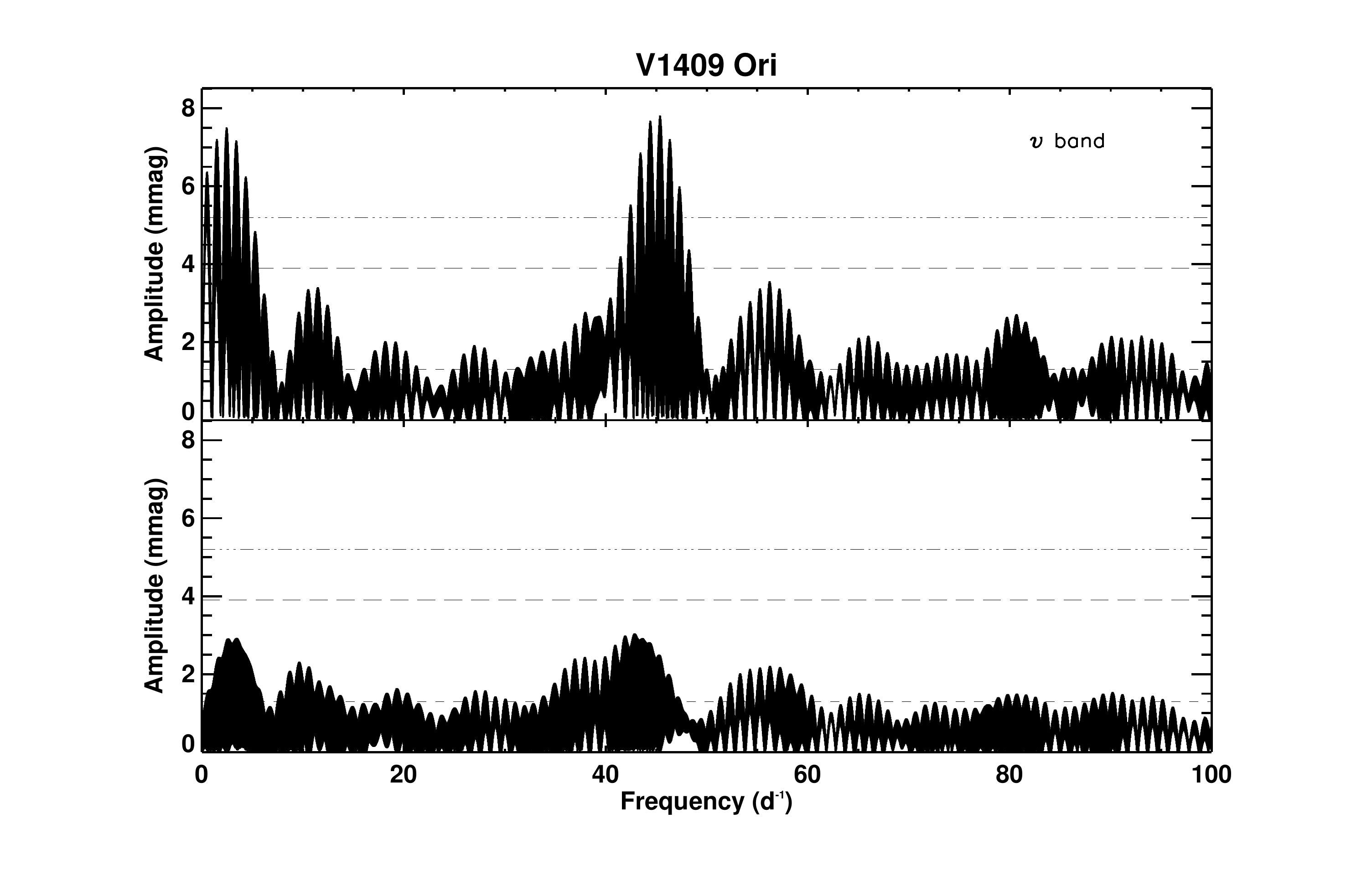}
  \includegraphics[width=9cm]{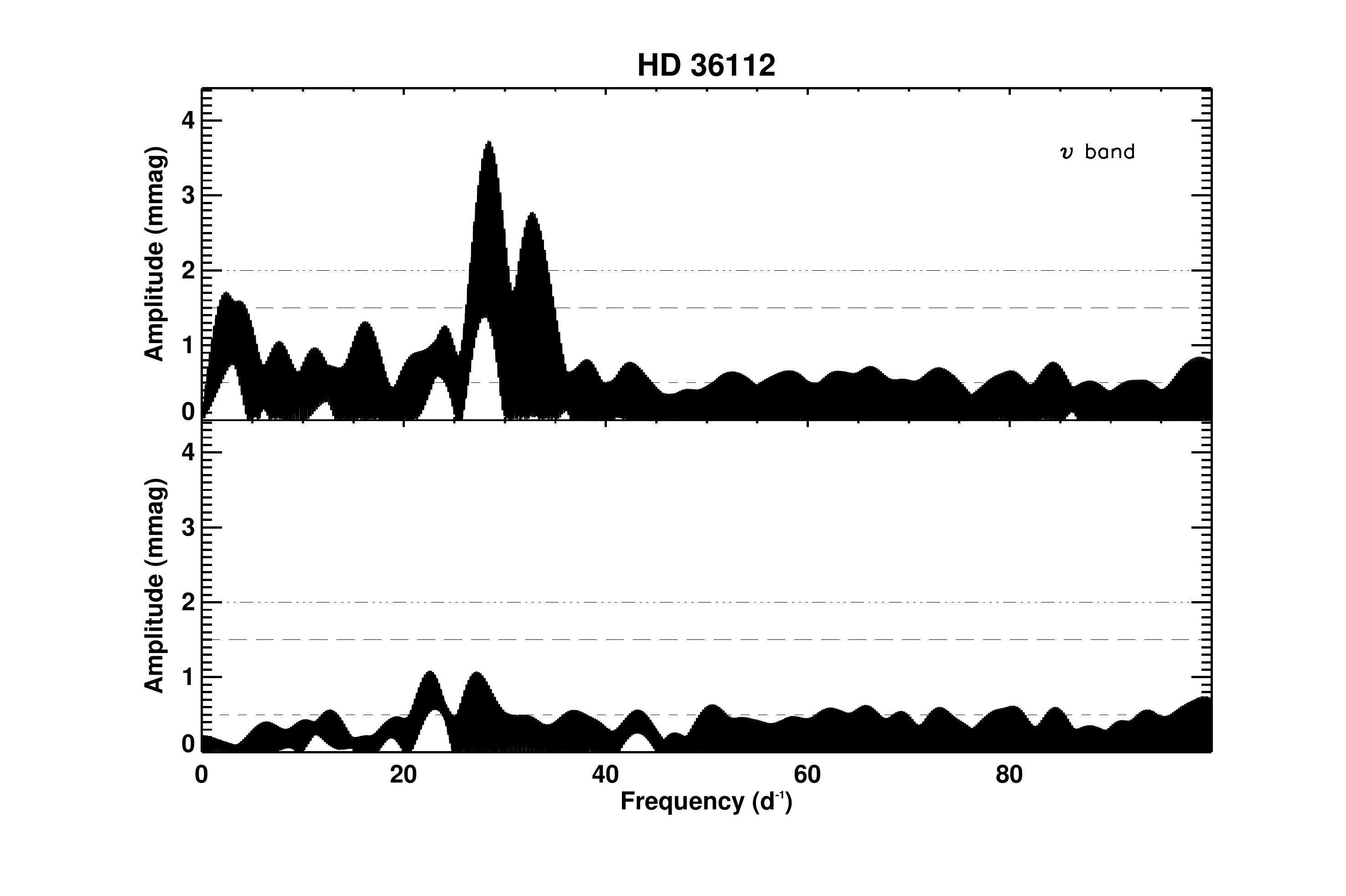}
   \includegraphics[width=9cm]{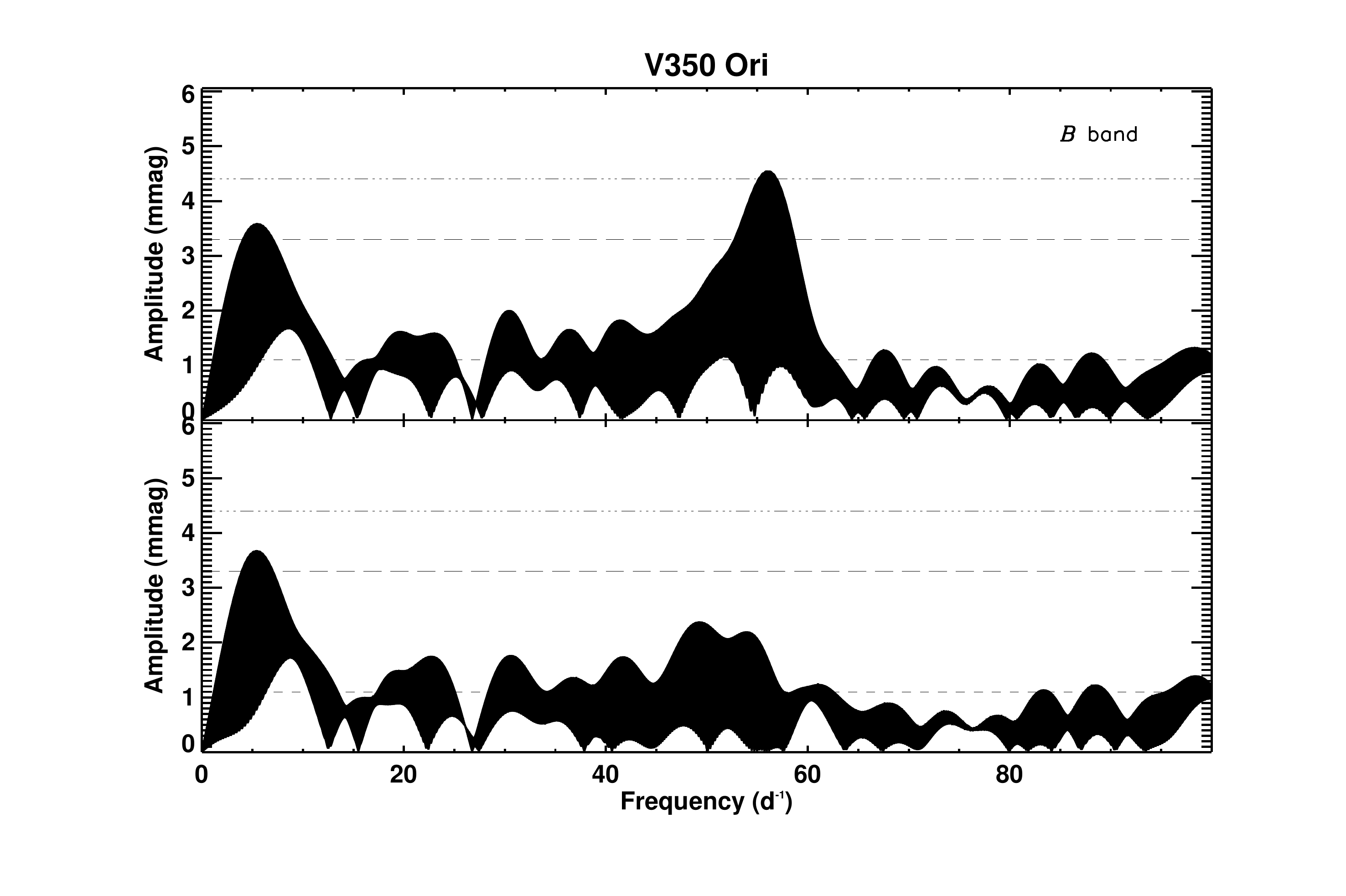}
   \includegraphics[width=9cm]{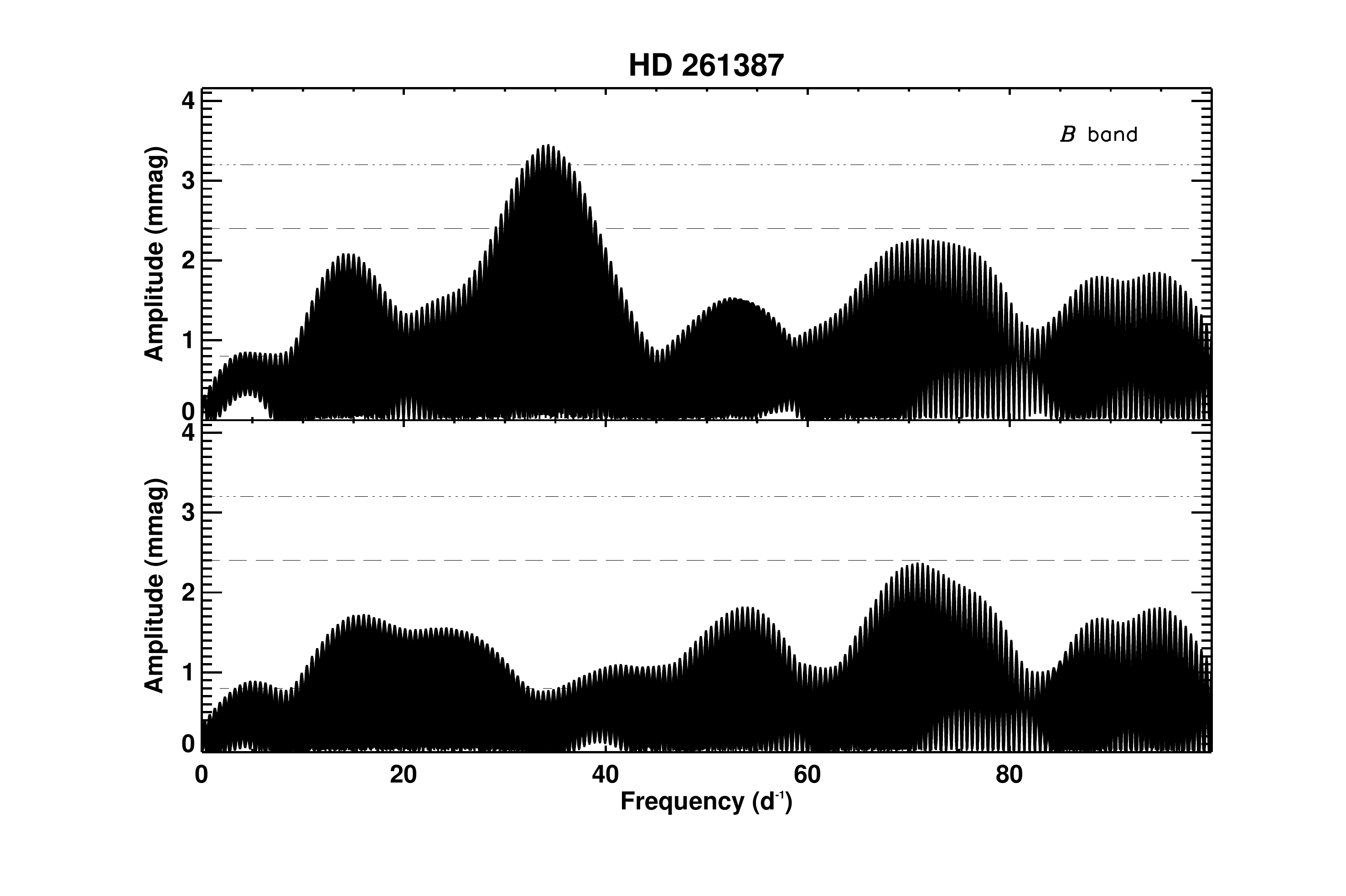}
   \includegraphics[width=9cm]{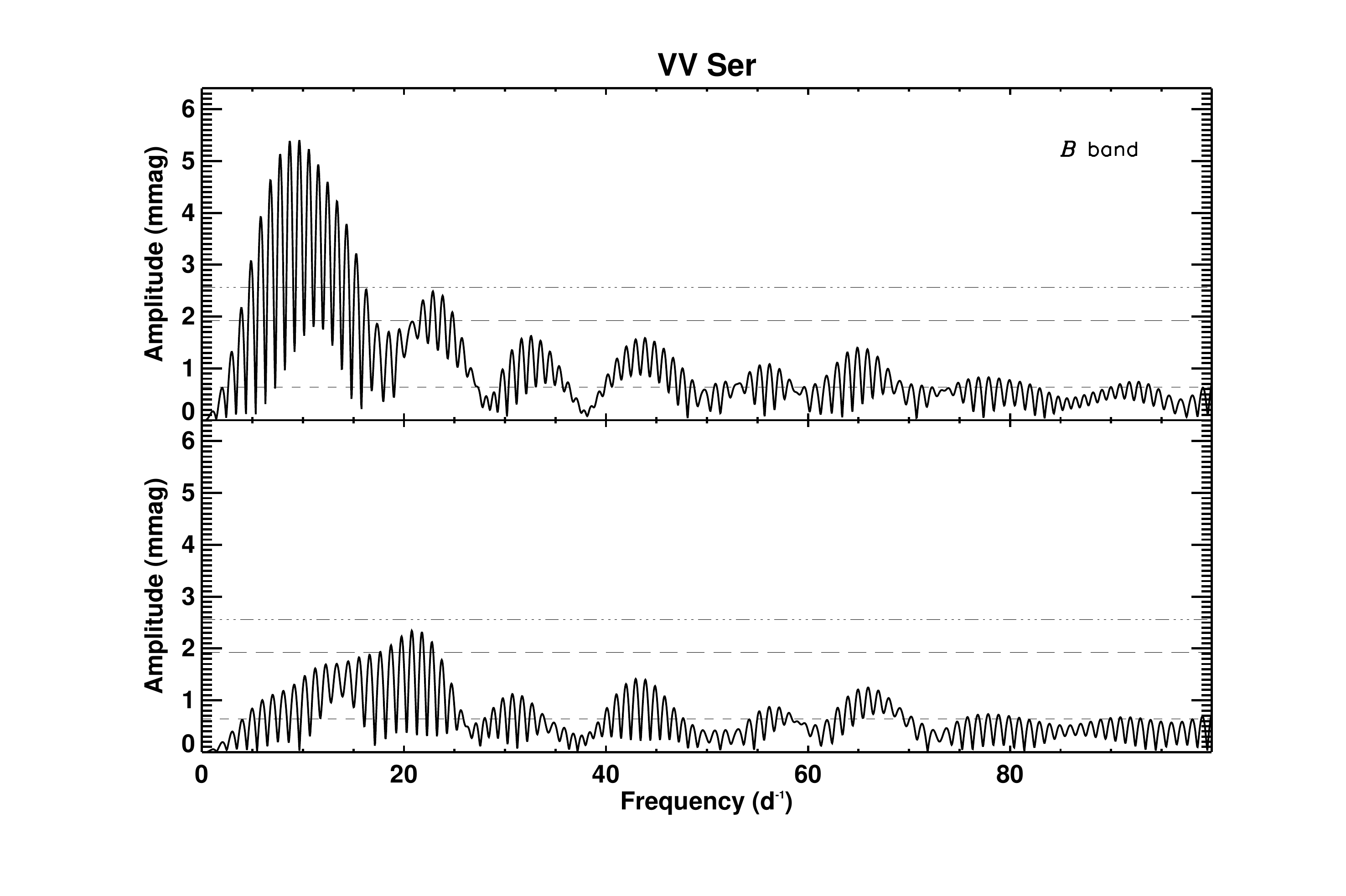}
      \caption{Frequency analysis in the \textit{v} or \textit{B} bands of the $\delta$ Scuti-type stars detected or confirmed in our sample. The dashed, long-dashed, and dot-dashed lines show the $\sigma$, 3$\sigma$, and 4$\sigma$ levels, respectively.
               \textit{Bottom panels}: Residuals after prewhitening the main frequency peak detected (see Table \ref{tab:PMScortoper}).
              }
         \label{FigDeltaScuti}
   \end{figure*}

We have found that five stars show $\delta$ Scuti-type pulsations (PDS 004, HD 35187, V1409 Ori, HD 36112, and V350 Ori). In addition, two other stars (HD 261387 and VV Ser) 
have also been confirmed as $\delta$ Scuti-type pulsators. The results are summarised in Table~\ref{tab:PMScortoper}. Frequency and amplitude errors are indicated in parentheses.
These stars fulfill the variability criterion in all the bands observed, whereas the rest of the stars that are listed in Tables~\ref{tab:StarsI} and \ref{tab:StarsII} do not. Therefore, it is 
assumed that the peaks found in these latter objects are not caused by real variability. Figure \ref{FigDeltaScuti} shows the results of the frequency analysis performed in the \textit{v} 
or \textit{B} bands for the PMS $\delta$ Scuti-type stars detected or confirmed. \\

\begin{figure*}
   \centering
   \includegraphics[width=9cm]{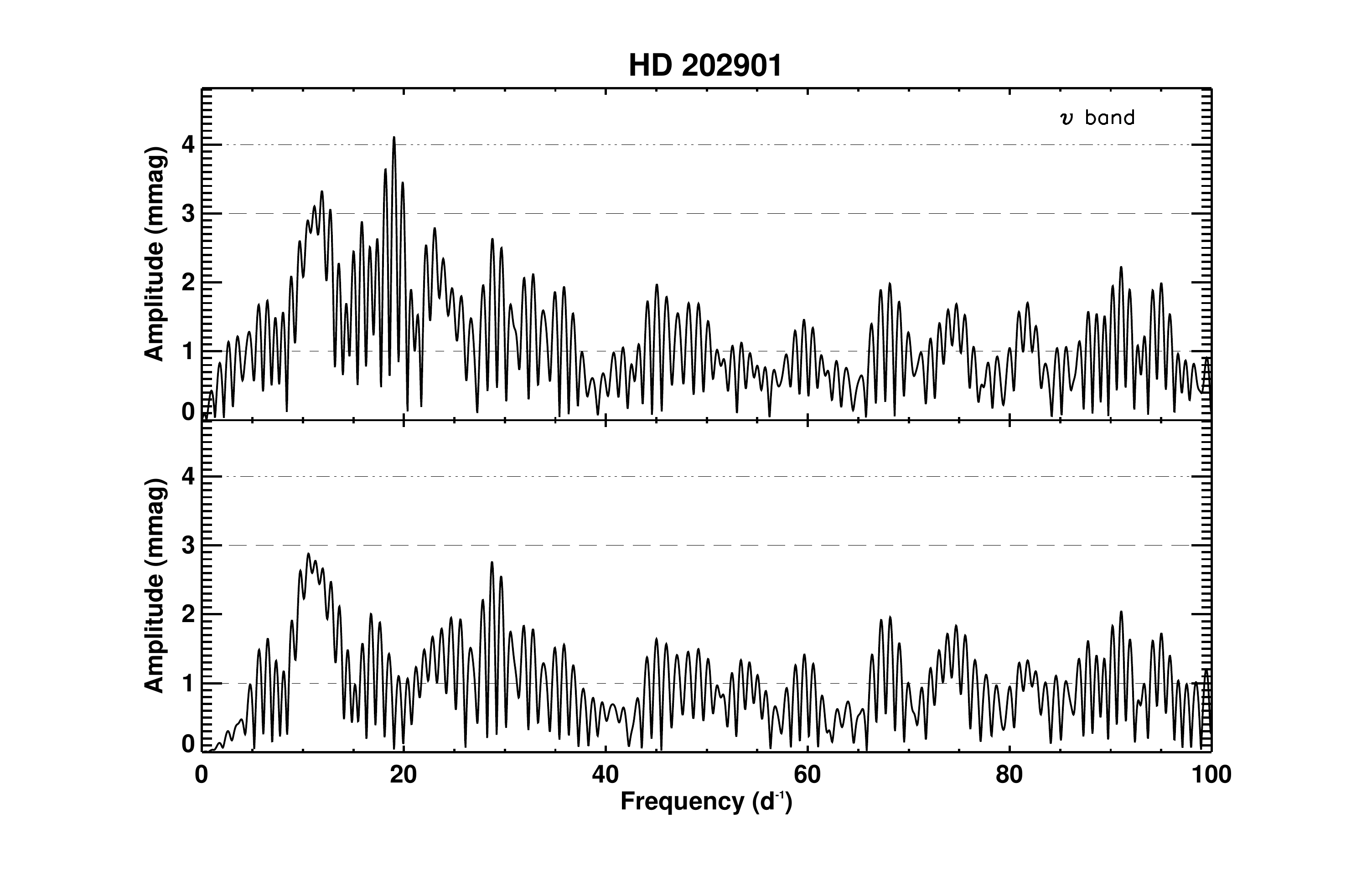}
  \includegraphics[width=9cm]{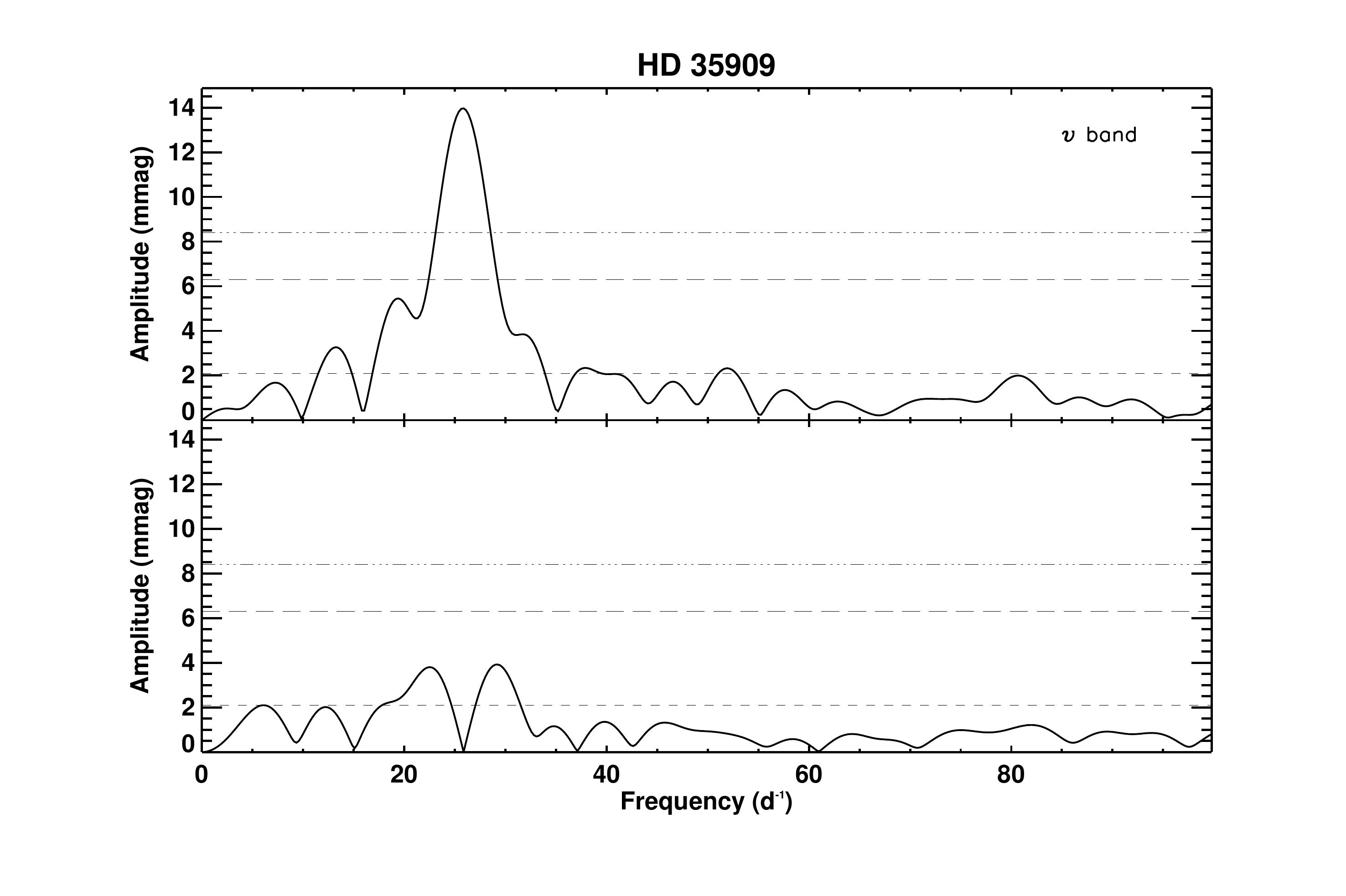}
   \includegraphics[width=9cm]{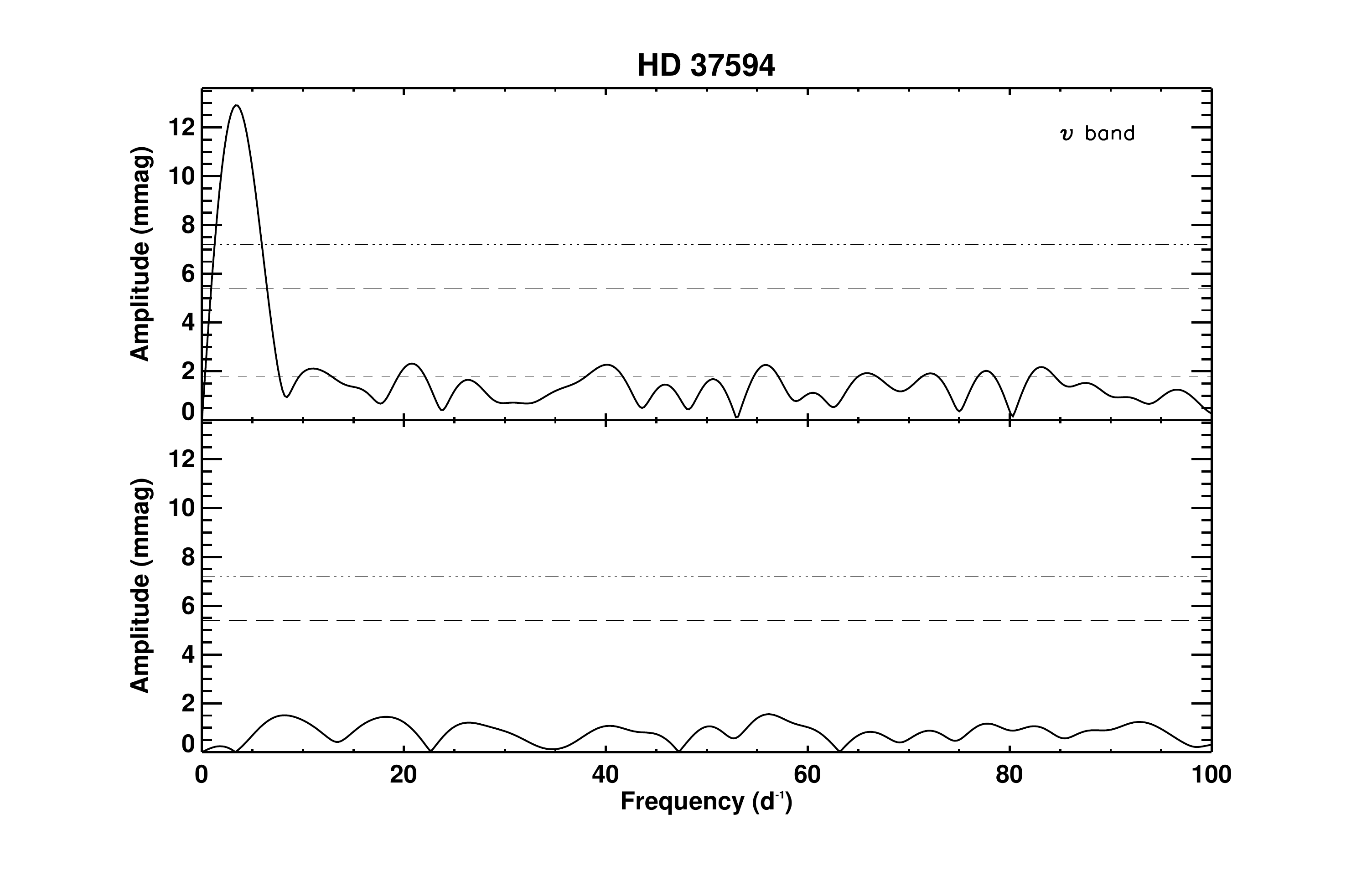}
  \includegraphics[width=9cm]{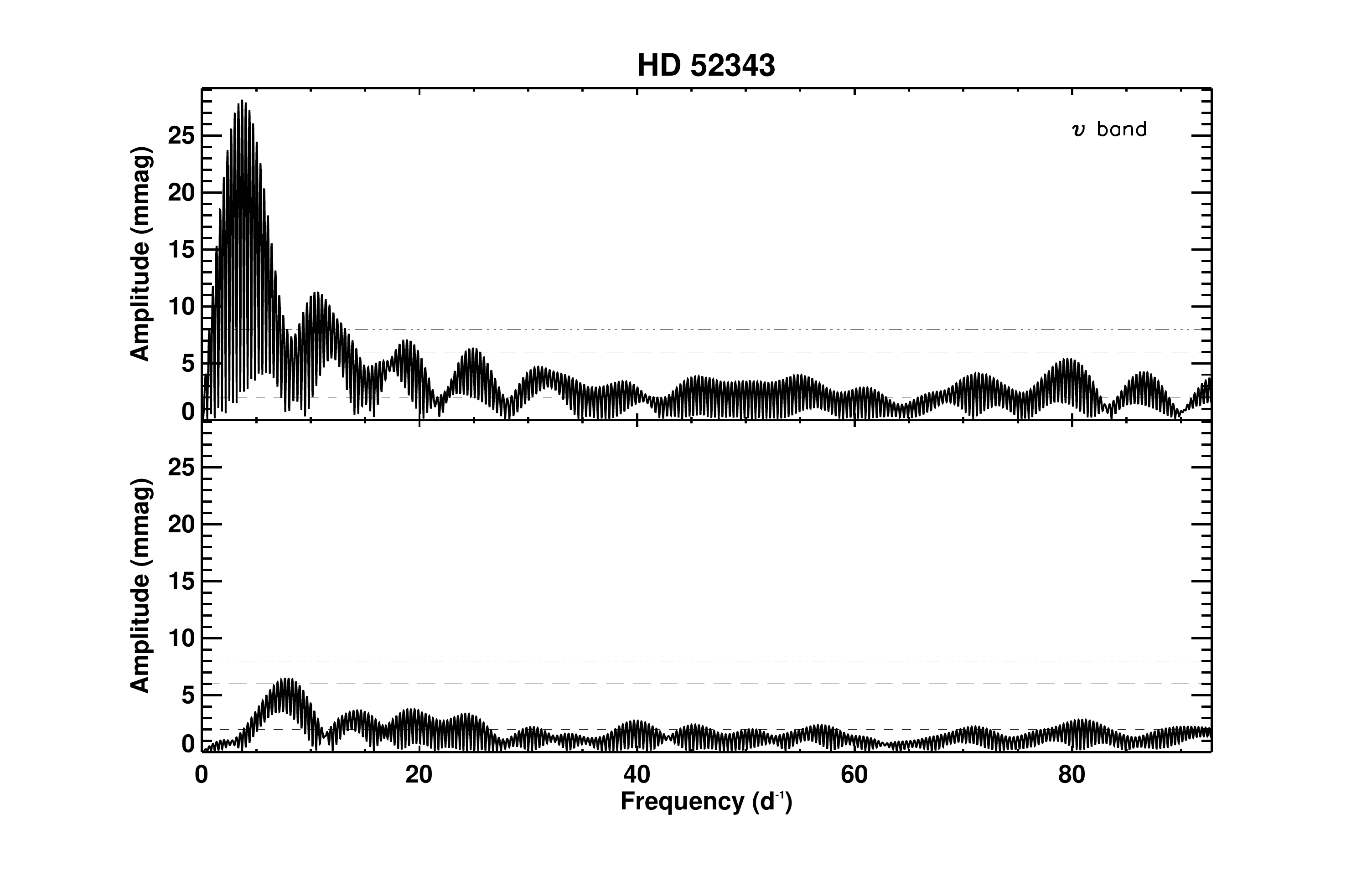}
   \includegraphics[width=9cm]{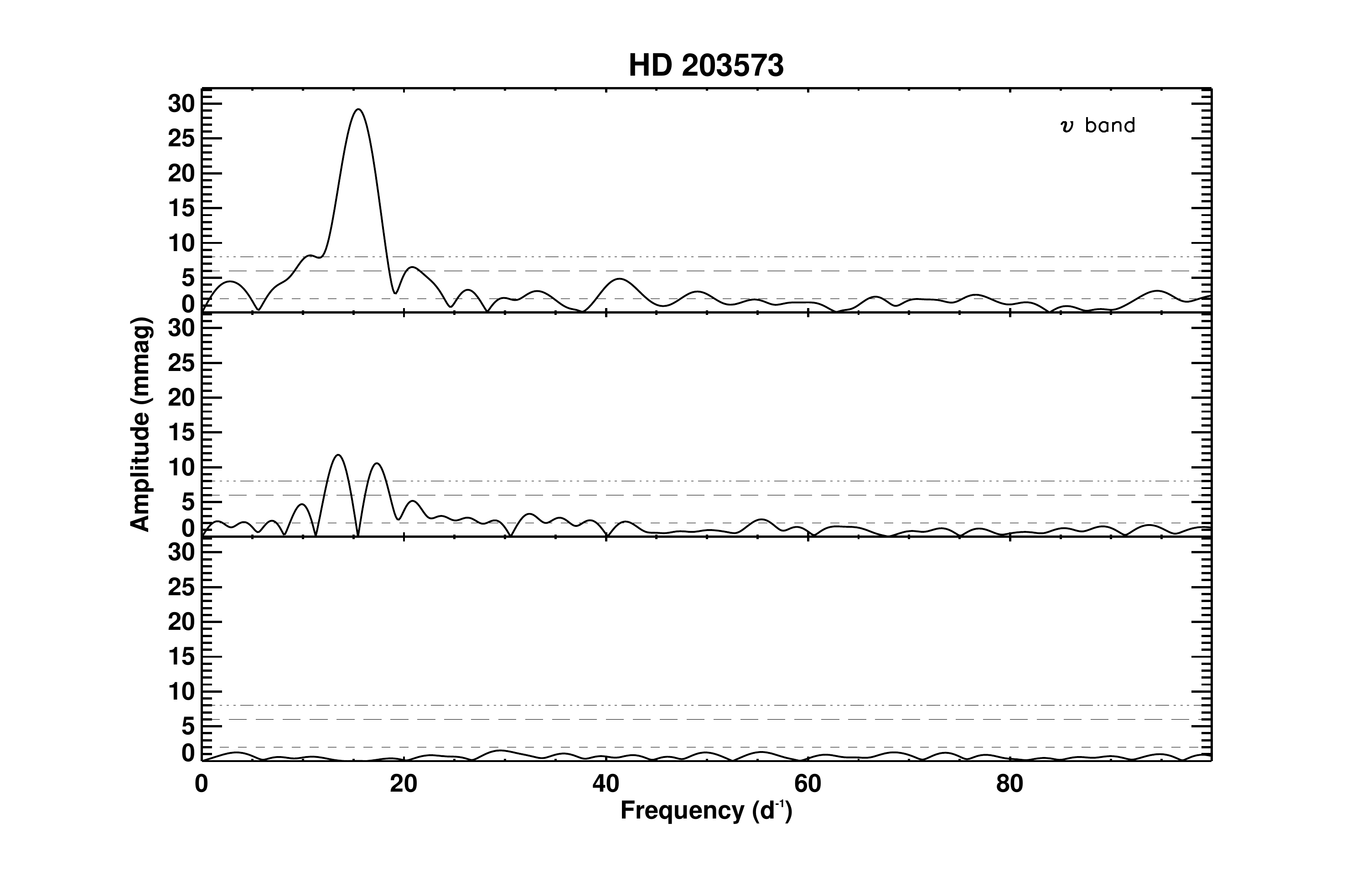}
   \caption{Frequency analysis in the \textit{v} band of the MS comparison and check stars in which $\delta$ Scuti or $\gamma$ Doradus-type pulsation have been detected. 
              The dashed, long-dashed, and dot-dashed lines show the $\sigma$, 3$\sigma$, and 4$\sigma$ levels, respectively.
               \textit{Bottom panels}: Residuals after prewhitening the main frequency peak detected (see Table \ref{tab:VarComp}).
              }
         \label{FigCompStars}
   \end{figure*}

On the other hand, three new MS $\delta$ Scuti- and two new MS $\gamma$ Doradus-type stars have also been detected among the comparison and check stars. After a few hours, changes in the 
luminosities of these stars were revealed as evident with probable pulsational periodicities. Thus, these stars were rejected 
as comparison or check stars and re-observed to confirm the periodicities. In all cases, they were discovered using the T90 and \textit{uvby} photometry. Figure \ref{FigCompStars} shows 
the respective amplitude spectra in the Strömgren \textit{v} band 
for the new variables. Table~\ref{tab:VarComp} shows the results obtained from the frequency analysis and some basic information of the observations carried out 
for each star. The last column indicates the ID numbers of the objects to which these comparison and check stars were initially associated.

\begin{table*}
\caption{PMS and HAEBE stars in which $\delta$ Scuti-type pulsation have been detected (ID 3,9,14,15, and 26) or confirmed (ID 63 and 42).} 
\centering
\scalebox{0.92}{
\begin{tabular}{cccccccccccc}  \hline
ID&Name& Spectral & Frequency & $\Delta$\textit{B} & $\Delta$\textit{V} & $\Delta$\textit{I} & $\Delta$\textit{u} & $\Delta$\textit{v} & $\Delta$\textit{b} & $\Delta$\textit{y} & Pulsation \\
      \cline{5-11}
number&~    & Type & (d$^{-1}$)  & \multicolumn{7}{c}{(mmag)} & \\
\hline
3 &PDS 004&A1&53.27 (1) & 10.7 (7) & 8.5 (7) &6.2 (7) & - & - & - & - & $\delta$ Scuti \\ 
9 &HD 35187&A2e&63.87 (1) & - & -  & -  & 3.6 (7) & 3.3 (4) & 3.1 (4) & 2.5 (4) & $\delta$ Scuti \\
14&V1409 Ori&AIab&45.35 (1) & - & - & - & 7.6 (11) & 7.8 (7) & 6.4 (5) & 5.4 (7) & $\delta$ Scuti \\
15&HD 36112&A3e&28.36 (1) & - & - & - & 3.0 (6) & 3.7 (3) & 3.5 (3) & 2.8 (3) & $\delta$ Scuti \\
26&V350 Ori&A0&57.08 (1) &4.5 (5)&3.3 (4)& 3.0 (4)& - & - & - & - & $\delta$ Scuti \\
63&HD 261387&A2V&34.67 (1) &3.5 (3)& 2.7 (3)& 2.3 (3) & - & - & - & - & $\delta$ Scuti \\
42&VV Ser&A2e&9.65 (4) & 5.8 (5) & 2.8 (4) & 2.6 (6) & - & - & - & - & $\delta$ Scuti\\
\hline
\end{tabular}
}
\label{tab:PMScortoper}
\end{table*}

\begin{table*}
\caption{Herbig Be stars in which probable $\beta$ Cephei-type pulsation have been detected.} %\vspace{3mm}
\centering
\scalebox{0.92}{
\begin{tabular}{cccccccccccc}  \hline
ID&Name& Spectral & Frequency & $\Delta$\textit{B} & $\Delta$\textit{V} & $\Delta$\textit{I} & $\Delta$\textit{u} & $\Delta$\textit{v} & $\Delta$\textit{b} & $\Delta$\textit{y} & Pulsation \\
      \cline{5-11}
number&~    & Type & (d$^{-1}$)  & \multicolumn{7}{c}{(mmag)} & \\
\hline
44&HD 174571&B1.5V& 4.51 (1) & - & - & - & 5.5 (8) & 4.7 (4) & 4.6 (4) & 3.5 (4) & $\beta$ Cephei? \\
50&V1685 Cyg&B3&7.10 (1) & 5.7 (6) & 4.0 (3) & 2.0 (4) & - & - & - & - & $\beta$ Cephei? \\
35&HD 50083&B2III& 6.45 (1) & - & - & - & 3.4 (5) & 2.7 (3) & 2.7 (3) & 2.7 (4) & $\beta$ Cephei? \\
\hline
\end{tabular}
}
\label{tab:Betacortoper}
\end{table*}

\begin{table*}
\caption{MS comparison and check stars in which $\delta$ Scuti or $\gamma$ Doradus-type pulsation have been detected.} %\hspace{100mm}
\centering
\scalebox{0.92}{
\begin{tabular}{cccccccccc}  \hline
Name     & V     &Spectral & Frequency    & $\Delta$\textit{v}   &  Season        & Hours          &  Points         & Pulsation      &  ID                 \\ 
  ~      & (mag) &Type     & (d$^{-1}$)   & (mmag)      &                & (nights)       &                 &                &  number             \\ 
\hline 
HD 202901& 8.350 & F0      & 19.00 (7)       & 4.0 (6)     & Aug09          & 11.45 (2)      &  94             & $\delta$ Scuti &  51,52              \\
HD 203573& 7.795 & F0      & 16.20 (3)       & 23.3 (4)    & Aug09          &  8.31 (1)      &  124            & $\delta$ Scuti &  51,52              \\
         &       &         & 14.41 (3)       & 19.8 (4)    &                &                &                 &                &                     \\
HD 35909 & 6.390 & A4      & 25.61 (10)      & 14.0 (5)    & Oct07          & 5.01 (1)       &  70             & $\delta$ Scuti &  58                 \\
HD 37594 & 5.990 & A8      & 2.80  (9)       & 14.3 (5)    &  Dec07         & 5.42 (1)       &  65             & $\gamma$ Dor   &  13,19,26           \\
HD 52343 & 8.358 & F0      & 3.69  (1)       & 26.0 (9)    &  Feb08         & 8.41 (2)       &  54             & $\gamma$ Dor   &  36,37,38,39,40,62  \\
\hline
\end{tabular}
}
\label{tab:VarComp}
\end{table*}

\subsection{PDS 004}

The object PDS 004 was observed at 2.21 and 3.96 hours on two different nights in November 2007 and 2008, respectively, with T150 in the \textit{BVI} bands. Additionally, it was observed at 4.62
 hours on only one night in December 2009 with T90 and Strömgren \textit{uvby} bands. It shows a main frequency peak at \textit{ f}$_{1}$ = 53.27 d$^{-1}$. This pulsation frequency was
also detected in the \textit{BI} bands with S/N = 4.2. Amplitude ratios between different bands ($\Delta$B = 10.7 mmag > $\Delta$V = 8.5 mmag > $\Delta$I > 
6.2 mmag and ${\Delta B \over \Delta V}$ = 1.25, ${\Delta I \over \Delta V}$ = 0.72) are in good agreement with those typical of $\delta$ Scuti-type pulsations 
\citep{1990A&A...234..262G,1996A&A...307..539R,1999MNRAS.302..349B,2005ASPC..333..165R}. In these papers, it can be seen how the amplitude ratios for light curves 
observed in different photometric bands in a pulsating star mainly depend on the variations in temperature and gravity during the pulsational 
cycles. This means that they mainly depend on the variations of these physical parameters in a colour-colour grid, such as (c$_{1}$,(\textit{b$-$y})) or 
(c$_{1}$, $\beta$) grids, but it is not expected to depend (at least not largely) on its evolutionary status as a MS or PMS star.

\subsection{HD 35187}

The star HD 35187 was observed from October-December 2007 (9.87 hours) and December 2009 (1.41 hours) using T90 and Strömgren \textit{uvby} bands. It shows a main frequency peak at 
\textit{ f}$_{1}$ = 63.87 d$^{-1}$ with a S/N = 4.7. The amplitude ratios between different \textit{vby} bands ($\Delta$v = 3.3 mmag > $\Delta$b = 3.1 mmag > $\Delta$y = 2.5 mmag and 
${\Delta b \over \Delta y}$ = 1.24, ${\Delta v \over \Delta y}$ = 1.32) agree well with a $\delta$ Scuti-type pulsation. On the other hand, \textit{ f}$_{2}$ = 57.45 d$^{-1}$ and 
\textit{ f}$_{3}$ = 6.00 d$^{-1}$ frequencies show peaks with amplitudes between the 3$\sigma$ and 4$\sigma$ significance levels. The frequency \textit{ f}$_{2}$ is detected in 2007 and 2009 data but \textit{ f}$_{3}$ 
appears only in 2007 data. This leads us to consider \textit{ f}$_{3}$ as only suspect. 

\subsection{V1409 Ori}

The object V1409 Ori was observed with T90 and Strömgren \textit{uvby} bands during four nights: October 2007 (4.94 hours), November 2007 (5.94 hours) and December 2009 (4.07 hours). It shows a short period pulsation
at \textit{ f}$_{1}$ = 45.35 d$^{-1}$. This frequency is observed in \textit{vby} bands with amplitudes larger than the significance level 4$\sigma$. In the case of the \textit{u} band, this peak centred at 45.35 d$^{-1}$ is also
detected but with an amplitude below the 3$\sigma$ significance level due to the high dispersion in the data at this particular band ($\Delta$u = 7.6$\pm$1.1 mmag). Amplitude ratios at this frequency 
also agree well with a $\delta$ Scuti-type pulsation ($\Delta$v = 7.8 mmag > $\Delta$b = 6.4 mmag > $\Delta$y = 5.4 mmag and ${\Delta b \over \Delta y}$ = 1.18, ${\Delta v \over \Delta y}$ = 1.44).

\subsection{HD 36112}

The star HD 36112 was observed with T90 and Strömgren \textit{uvby} bands for four nights (9.70 hours between October and December 2007 and 8.32 hours in December 2009). It shows two peaks 
above the 4$\sigma$ level at frequencies \textit{ f}$_{1}$ = 28.36 d$^{-1}$ and \textit{ f}$_{2}$ = 33.00 d$^{-1}$. These frequencies were detected in all \textit{uvby} Strömgren bands. Amplitude 
ratios between bands at  \textit{ f}$_{1}$ are characteristic of a $\delta$ Scuti-type pulsation ($\Delta$v = 3.7 mmag > $\Delta$b = 3.5 mmag > $\Delta$y = 2.8 mmag $\simeq$ $\Delta$u = 3.0 mmag 
and ${\Delta b \over \Delta y}$ = 1.25, ${\Delta v \over \Delta y}$ = 1.32). After prewhitening of \textit{ f}$_{1}$ and \textit{ f}$_{2}$, residuals show two peaks at \textit{ f}$_{3}$ = 2.62 d$^{-1}$ 
and \textit{ f}$_{4}$ = 3.22 d$^{-1}$ with S/N > 3.0. These peaks are most probably generated by a gradient between the two observing nights on October 4 and 9, 2007, which are caused by a bad extinction correction.

\subsection{V350 Ori}

The object V350 Ori was observed using T150 in the Johnson-Cousins \textit{BVI} bands during four nights, two in January 2008 (5.24 hours) and another two in November 2008 (2.23 hours). It shows a 
main frequency peak located at \textit{ f}$_{1}$ = 57.08 d$^{-1}$ with S/N = 3.7. As mentioned in Sect. 3, the \textit{TRIPP} software showed this peak with a probability $>$ 99$\%$, suggesting it
as a real peak. The same peak at \textit{ f}$_{1}$ = 57.08 d$^{-1}$ is also observed in \textit{BI} bands with S/N = 3.9 and S/N = 3.7, respectively, with a probability > 99$\%$. 
All this suggests that the peak \textit{ f}$_{1}$ is real. Moreover, the amplitudes corresponding to different bands are in good agreement with $\delta$ Scuti-type pulsations 
($\Delta$B = 4.5 mmag > $\Delta$V = 3.3 mmag > $\Delta$I = 3.0 mmag and ${\Delta B \over \Delta V}$ = 1.36, ${\Delta I \over \Delta V}$ = 0.9).

\subsection{HD 261387}

The star HD 261387 has already proved to be a $\delta$ Scuti-type pulsator by observations carried out using the \textit{MOST} satellite \citep{2003PASP..115.1023W} between 
December 2006 and January 2007 and has been published by \citet{2009A&A...502..239Z}. The star was included in our study to confirm these findings and support them with new observations.  
It was observed with both T90 and T150 telescopes and the Strömgren \textit{uvby} and Johnson \textit{BVI} bands, respectively. In the first case, it was observed for
only one night in December 2007 for 6.87 hours. In the second case, it was observed during four different nights: two nights in January 2008 for 6.14 hours, one night in November 2008 for 2.98 hours, and one
last night in April 2010 for 2.40 hours. Frequency analysis shows a peak located
at \textit{ f}$_{1}$ = 34.67 d$^{-1}$ with amplitude $\Delta$V = 2.7 mmag and S/N = 4.5. The same peak is observed in the \textit{B} band with an amplitude $\Delta$B = 3.5 mmag and S/N = 4.0, 
which fulfills the amplitude ratio ${\Delta B \over \Delta V}$ = 1.29 and agrees well with a $\delta$ Scuti-type pulsation. In the case of Johnson \textit{I} band, we have found 
a peak at the same frequency \textit{ f}$_{1}$ = 34.67 d$^{-1}$, which is not significant (< 3$\sigma$ level) due to the higher noise level present in the periodograms for this band. 

\subsection{VV Ser}

The star VV Ser was observed using T150 and \textit{BVI} bands for three nights in April 2007 for a total observing time of 7.01 hours and for another night in April 2010 for 1.75 hours.
For the first time, \citet{2007A&A...462.1023R} detected that VV Ser pulsates as a $\delta$ Scuti through observations carried out for three consecutive years (2002-2004). For the best data set, 
which corresponds to 2004, these authors detected seven different pulsation frequencies with values between 2.69 and 10.24 d$^{-1}$. In our case, we have detected a main frequency peak at  
\textit{ f}$_{1}$ = 9.65 d$^{-1}$ that is very probably related to the frequencies \textit{ f}$_{5}$ = 9.55 d$^{-1}$ (2002 data set), \textit{ f}$_{5}$ = 8.50 d$^{-1}$ (2003 data set), and 
\textit{ f}$_{6}$ = 7.56 d$^{-1}$ (2004 data set), which are detected by \cite{2007A&A...462.1023R}. Therefore, we can confirm VV Ser as a $\delta$ Scuti-type pulsator.

\section{$\beta$ Cephei stars} 

\begin{figure}
   \centering
   \includegraphics[width=\hsize]{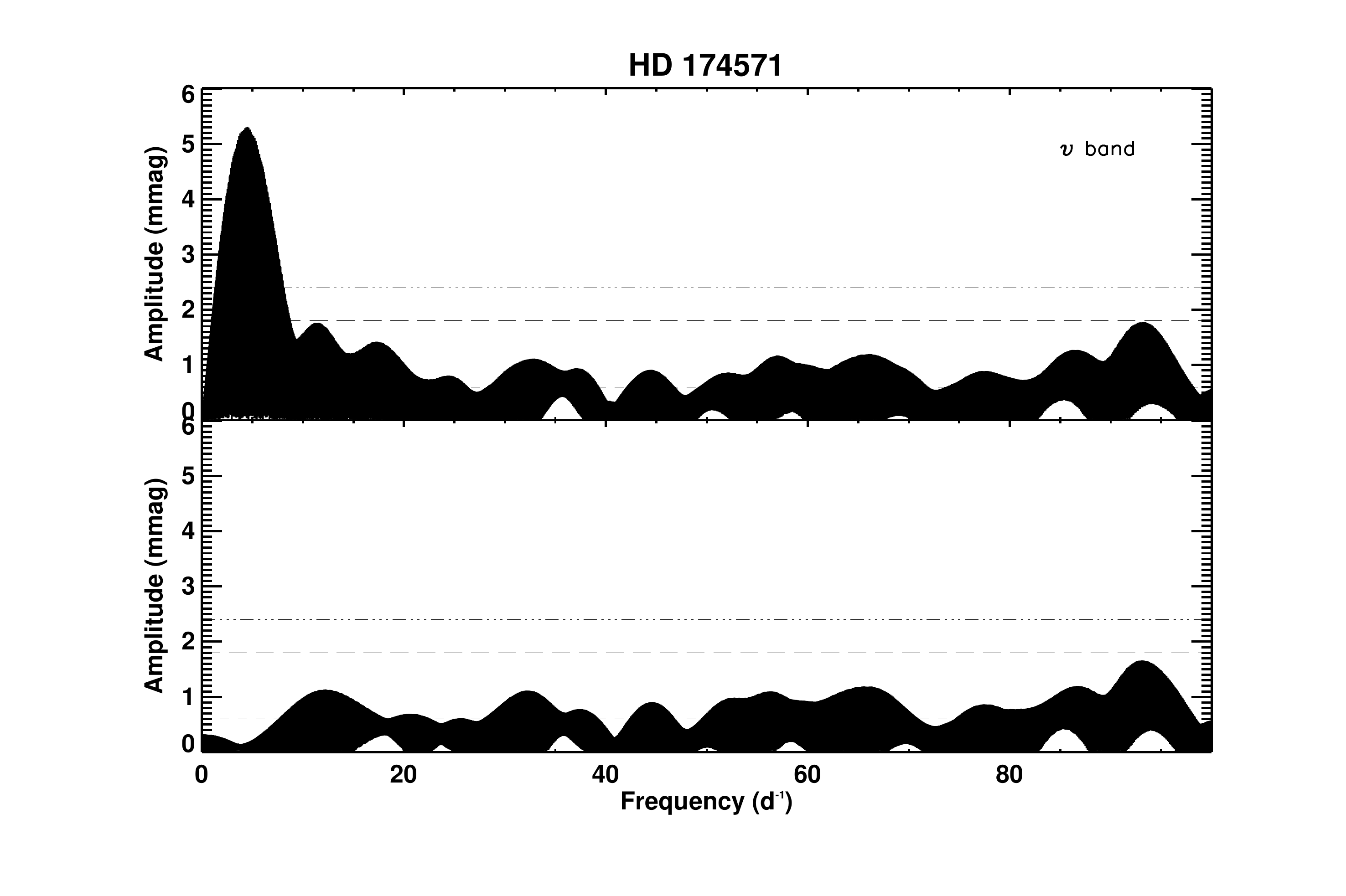}
   \includegraphics[width=\hsize]{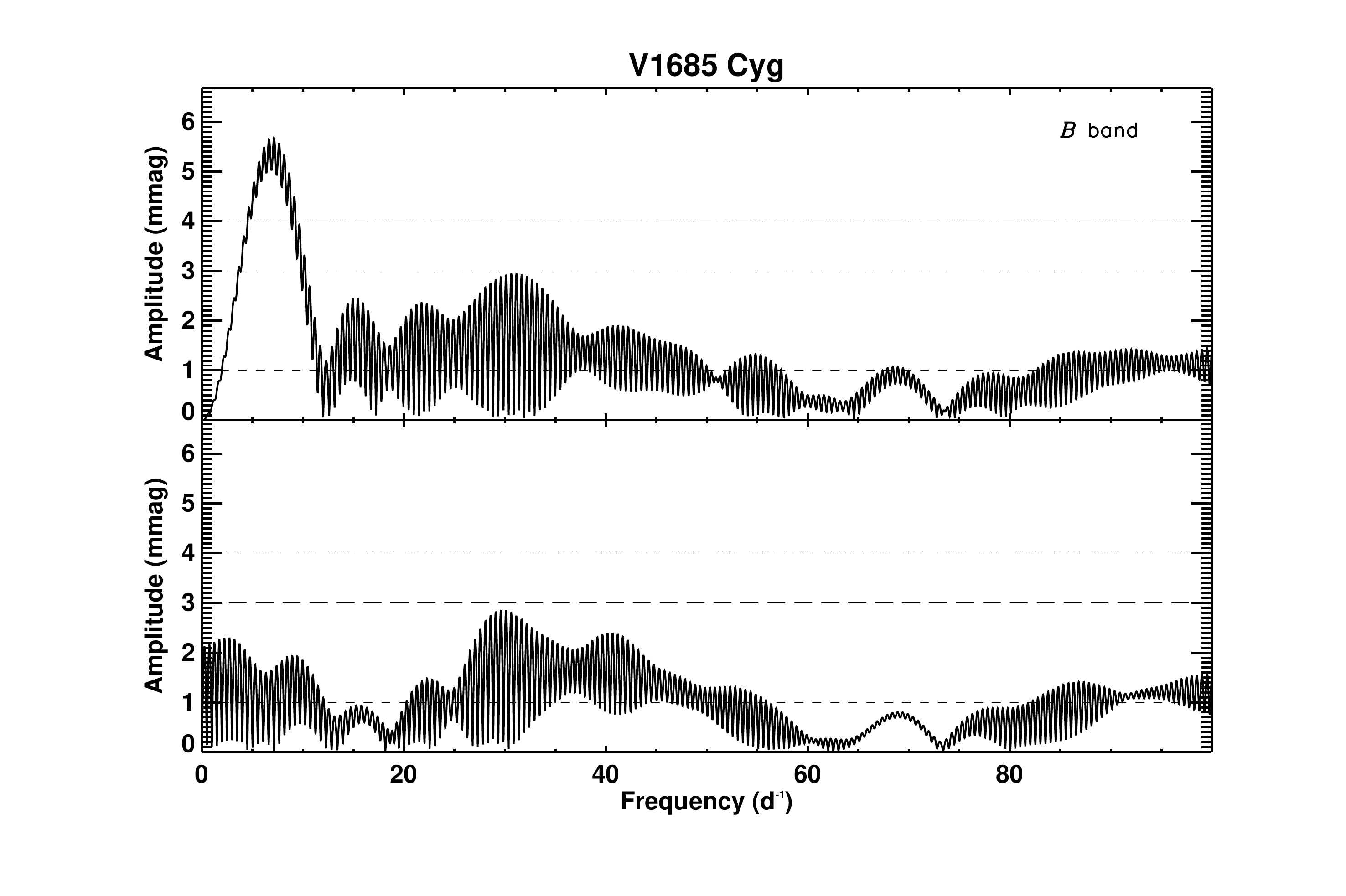}
   \includegraphics[width=\hsize]{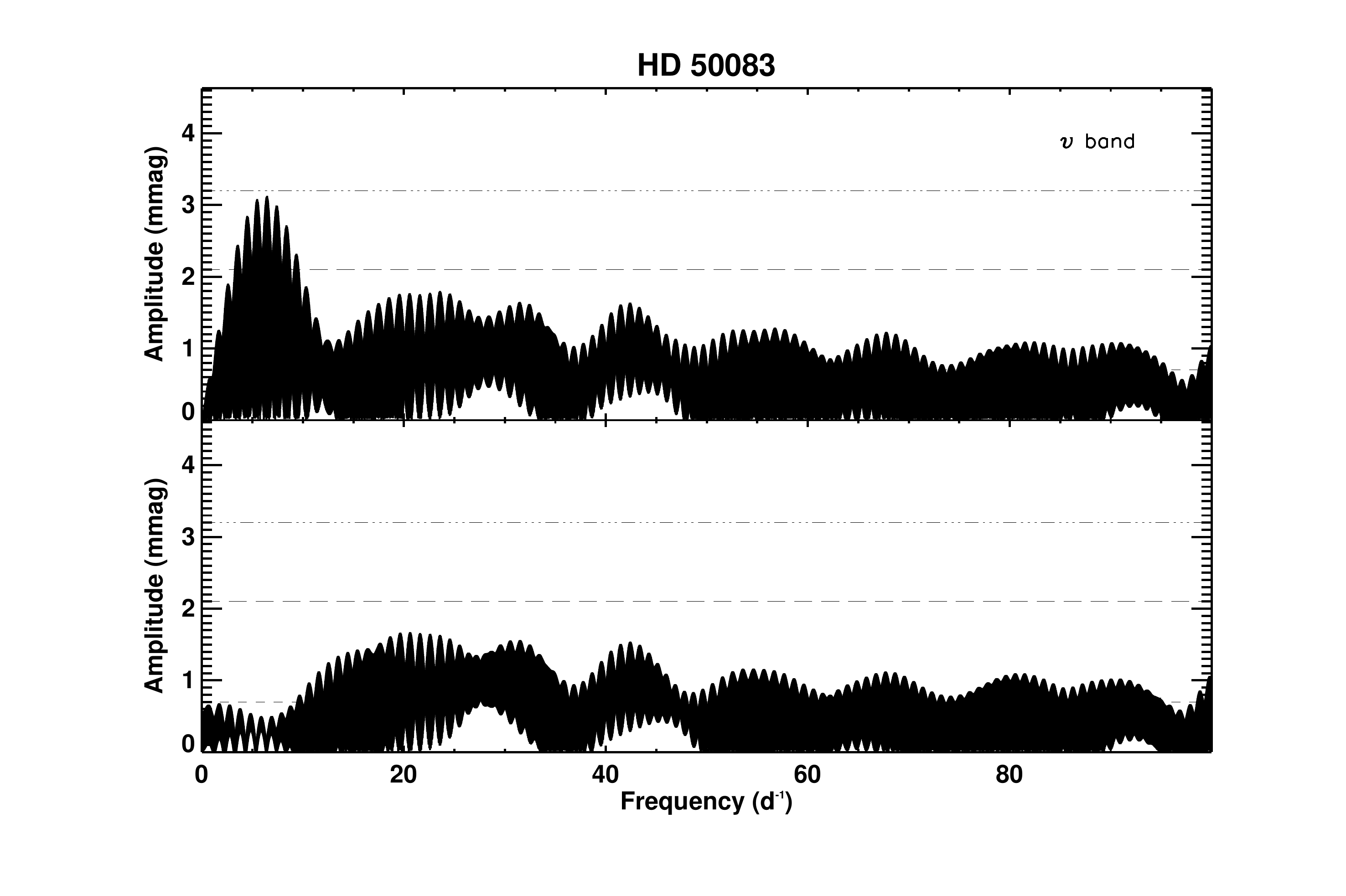}
      \caption{Frequency analysis in the \textit{v} or \textit{B} bands of the $\beta$ Cephei-type stars detected. The dashed, long-dashed, and dot-dashed lines show the $\sigma$, 3$\sigma$, and 4$\sigma$ levels, respectively. 
               \textit{Bottom panels}: Residuals after prewhitening the main frequency peak detected (see Table \ref{tab:Betacortoper}).
              }
         \label{FigBetaCephei}
   \end{figure}

Three stars are probably $\beta$ Cephei-type pulsators (HD 174571, V1658 Cyg, and HD 50083). The results are summarised in Table~\ref{tab:Betacortoper}.
Frequency and amplitude errors are indicated in the parentheses. Figure \ref{FigBetaCephei} shows the results of the frequency analysis performed 
in the \textit{v} or \textit{B} bands for these stars. It is important to note that stars more massive than about 8 M$_{\odot}$ do not have an optically 
visible PMS phase, as the birthline intersects the ZAMS at about this mass \citep{1990ApJ...360L..47P}. The objects HD 174571, V1658 Cyg, and HD 50083 are classified as
early B stars (M > 8M$_{\odot}$); hence, they are probably young MS stars. On the other hand, it is necessary to obtain more observations to confirm their $\beta$ Cephei nature.

\subsection{HD 174571}

The object HD 174571 is a B1.5V star \citep{2006A&A...451.1053F}. This star was observed on four different nights (May to July 2007, August 2009, and April 2010) with T90 and Strömgren \textit{uvby} bands 
for a total time of 14.12 hours. It shows a main frequency peak located at \textit{ f}$_{1}$ = 4.51 d$^{-1}$ in all Strömgren bands with amplitudes larger than the 4$\sigma$ level.
Amplitudes in different bands ($\Delta$u = 5.5 mmag > $\Delta$v = 4.7 mmag > $\Delta$b = 4.6 mmag > $\Delta$y = 3.5 mmag) are consistent with a $\beta$ Cephei-type pulsation 
\citep{2005ASPC..333..165R}. 
 
\subsection{V1685 Cyg}

The object V1685 Cyg is a B3 star \citep{2006ApJ...653..657M,2004AJ....127.1682H}. With T150, it was observed in the Johnson-Cousins \textit{BVI} bands for one night in August 2007 
(4.83 hours) and two nights in August 2009 for a total time of 5.24 hours. It shows a main frequency peak located at \textit{ f}$_{1}$ = 7.10 d$^{-1}$ in
all Johnson-Cousins \textit{BVI} bands. Amplitudes in different bands ($\Delta$B = 5.7 mmag > $\Delta$V = 4.0 mmag > 
$\Delta$I = 2.0 mmag) suggest that the variations detected are also caused by $\beta$ Cephei-type pulsations, which agrees well with its spectral type B3 and main pulsation frequency.

\subsection{HD 50083}

The star HD 50083 was observed using the T90 and Strömgren \textit{uvby} bands during two nights in December 2007 (5.36 hours) for one night in February 2010 (2.85 hours) and another one in February 2010 
(3.08 hours). It shows a main frequency peak located at \textit{ f}$_{1}$ = 6.45 d$^{-1}$ for the \textit{v} band with S/N = 3.8 and a probability >99$\%$,
suggesting this as a real peak. Moreover, this peak is present in all Strömgren bands with amplitudes $\Delta$u = 3.4 mmag > $\Delta$v, $\Delta$b, $\Delta$y = 2.7 mmag, and 
${\Delta v \over \Delta u}$ = 0.79, which suggests $\beta$ Cephei-type pulsations, according to its spectral type B2III \citep{2006A&A...451.1053F} and pulsation frequency.

\section{Stellar physical parameters}

Stellar physical parameters were determined using the colour and $\beta$ indices from Str\"{o}mgren photometry collected in this work. To obtain the dereddened indices 
(\textit{b$-$y})$_{0}$, \textit{m}$_{0}$, and \textit{c}$_{0}$, an initial classification of the stars in their different spectral groups was first necessary. For this purpose, 
the spectral classification given by \citet{1966ARA&A...4..433S} was used. Once the stars were classified, photometric indices were dereddened by following the equations and ZAMS 
reference lines derived by \citet{1980A&AS...40..199P}. 
Stars previously classified in the literature as Be stars were dereddened using the method developed by \citet{1998A&A...332..643F}. 
Table~\ref{tab:Dereddenedindices} shows the spectral type group; visual magnitude \textit{V}; photometric indices (\textit{b$-$y}), \textit{m}$_{1}$, and \textit{c}$_{1}$; and the dereddened indices
(\textit{b$-$y})$_{0}$, \textit{m}$_{0}$, \textit{c}$_{0}$, $\delta$\textit{m}$_{1}$, and $\delta$\textit{c}$_{1}$. The last two columns show the $\beta$ values, and the modified $\beta^{*}$ 
values obtained using the method developed by \citet{1998A&A...332..643F} in the case of Be stars. Typical error bars are about 0.$^{m}$01 for E(\textit{b$-$y}), (\textit{b$-$y})$_{0}$, \textit{m}$_{0}$, \textit{c}$_{0}$, $\delta$\textit{m}$_{1}$, and 0.$^{m}$02 for 
$\delta$\textit{c}$_{1}$. In the particular case of Be stars, typical error bars are about 0.$^{m}$04 for all indices. For 33 stars of the sample, there was
no \textit{uvby$\beta$} photometry available in the literature. For all the remaining stars, the \textit{uvby$\beta$} indices are listed in the catalogue of \citet{1998A&AS..129..431H}. Their 
values are in good agreement within the error bars with our results. \\

\begin{table*}
\caption{Dereddened photometric indices for the sample of observed stars. \textbf{Notes:} (1) High luminosity, $\delta$c$_{1}$ $\geq$ 0.$^{m}$28, (2) low metallicity subluminous star,
(3) Ap star with E(\textit{b-y})$\leq$-0.$^{m}$04 (derived value was E(\textit{b-y})=-0.$^{m}$065, adopted value is E(\textit{b-y})=0.$^{m}$000), (4) out of calibration range 
\citep{1980A&AS...40..199P}, (5) \textit{uvby$\beta$} photometry was not collected for this star, (6) derived value was E(\textit{b-y})=-0.$^{m}$028, adopted value E(\textit{b-y})=0.$^{m}$000, 
(7) Am star, m$_{0}$>0.$^{m}$22. } 
\centering
\scalebox{0.80}{
\begin{tabular}{cccccccccccccccc}
\hline \hline
ID&Name& Spectral& V &(\textit{b-y})& m$_{1}$& c$_{1}$&E(b-y) & (b-y)$_{0}$ &m$_{0}$&c$_{0}$&$\delta$m$_{1}$&$\delta$c$_{1}$&$\beta$&$\beta^{*}$&Note \\
number&    & type group& (mag)&  (mag)       & (mag)  & (mag)  &(mag) & (mag)       & (mag) & (mag) &   (mag)        &   (mag)       & (mag) & (mag)     &    \\
\hline
\\
1&VX Cas& B   &11.262&0.206 &0.068&1.130&0.217&-0.011&0.137&1.087&0.045 & ... & 2.913& ... & \\
2&V594 Cas&  B &10.983 &0.464&-0.057&0.258&0.587&-0.123&0.126&-0.073& -0.047 & ... &2.469& 2.600&  \\
3&PDS 004&  A-F &10.930 &0.241&0.144&0.948&0.144&0.097&0.191&0.919& 0.017 & 0.073 &2.839& ... & \\
4&XY Per& A0-A3 &9.486 &0.340&0.122&1.238&0.275&0.065&0.210&1.183&-0.002 &0.315&2.850& ... & 1 \\
5&AB Aur& A0-A3 &6.740 &0.108&0.120&1.003&0.051&0.057&0.136&0.993&0.069&0.191 &2.811& ...  & \\
6&HD 31648& A-F &7.384 &0.117&0.178&0.944&0.010&0.107&0.181&0.942&0.026 &0.118&2.825& ... & \\
9&HD 35187& A-F &7.889  &0.171&0.157&0.830&0.021&0.150&0.164&0.826&0.035 &0.074 &2.786& ... & \\
10&HD 287823& A0-A3 &9.867&0.092&0.171&0.899&0.036&0.056&0.182&0.892&0.025 &-0.028 & 2.875& ... & \\
12&HD 290409& A0-A3 &10.069&0.055&0.166&1.015&0.048&0.007&0.182&1.006&0.015 &0.026 &2.919& ... &\\
13&HD 290500& A-F &11.052&0.227&0.121&0.111&0.750&0.116&0.157&0.728&0.051 &-0.118 &2.838& ... & 2 \\
14&V1409 Ori& A-F &10.201 &0.118&0.192&0.870&0.000&0.118&0.192&0.870&0.010 &0.240 &2.736& ... & 3  \\
15&HD 36112& A-F  &8.466 &0.180&0.181&0.880&0.058&0.122&0.200&0.868&0.006 &0.057 &2.816& ... & \\
16&V1410 Ori& A-F &9.502  &0.099&0.193&0.986&0.016&0.083&0.198&0.983&0.010 &0.117 &2.849& ... & \\
17&HD 36408&  B &5.379 &0.046&0.075&0.762&0.090&-0.044&0.104&0.744&0.012 & ... & 2.704& ... & \\
18&V1271 Ori& A0-A3 &9.793&0.056&0.135&0.839&0.007&0.049&0.137&0.838&0.047 &0.159 &2.749& ... & \\
19&V380 Ori&  G &10.463 &0.365& 0.030&0.282& ... & ... & ... & ... & ... & ... &2.359& ... & 4  \\
20&HD 290770&  B  &9.196 &0.028&0.116&0.869&0.061&-0.033&0.135&0.857&-0.005 & ... &2.741& ... & \\
21&BF Ori&  A0-A3  &11.072 &0.267&0.157&1.221&0.232&0.035&0.231&1.175&-0.026 &0.233&2.892& ... & \\
22&HD 37357& A0-A3 &8.858 &0.069&0.169&0.975&0.022&0.047&0.176&0.971&0.031 &0.065 &2.868& ... &\\
23&V1247 Ori&  A-F &10.176 &0.193&0.213&0.831&0.000&0.193&0.213&0.831&-0.029 &0.273 &2.712& ... & 6  \\
24&V1788 Ori& A0-A3 &9.815& 0.102&0.164&1.009&0.043&0.059&0.178&1.000&0.029 &0.068 &2.884& ... & \\
25&RR Tau&  ...  & ... & ... & ... & ... & ... & ... & ... & ... & ... & ... & ... & ... & 5  \\
26&V350 Ori& A0-A3 & 10.879 &0.149&0.247&1.169&0.108&0.041&0.282&1.147&-0.079 &0.196 &2.900& ... & 7 \\
27&HD 37806&  B  &7.650 &0.059 &0.086&0.838&0.096&-0.037&0.117&0.819&0.003 & ... &2.713& ... &  \\
28&HD 38120&  B  &8.810 &0.044&0.119&0.949&0.069&-0.025&0.141&0.935&-0.030 & ... &2.692& ... & \\
30&HD 249879&  B  &10.253 &0.041 &0.122&0.876&0.073&-0.032&0.146&0.861&-0.004 & ... &2.775& ... & \\
31&V791 Mon&  B  &10.429 &0.242 &0.013&0.320&0.357&-0.115&0.122&0.009&-0.034 & ... &2.462& 2.620 & \\
32&HD 250550&  B &9.175 &0.082 &0.085&0.624&0.140&-0.058&0.130&0.596&-0.035 &... &2.654& ... & \\
33&AE Lep& A-F & 10.006 & 0.103&0.118&0.925&0.008&0.095&0.121&0.923&0.087 &0.070 &2.842& ... & \\
34&PDS 126& A0-A3 &12.089 &0.321&0.126&0.954&0.271&0.050&0.213&0.900&-0.009 &-0.050 &2.899& ... & \\
35&HD 50083&  B  &6.958 &0.104 &-0.002&-0.039&0.248&-0.144&0.073&-0.293& ... & ... &2.440& 2.530 & 4  \\
36&HD 52721&   B  &6.369 &0.098&0.029&0.001&0.225&-0.127&0.099&-0.113&0.058 & ... &2.562 & 2.589 & \\
37&HT Cma& B & 11.819 &0.327&0.022&1.033&0.350&-0.023 &0.134 &0.963 &0.010 & ... &2.796& ... &  \\
38&HU Cma&  B  &12.161 &0.255 &-0.010&0.647&0.306&-0.081&0.092&0.359&0.022 & ... &2.514& 2.700 &  \\
39&HD 53367& B &7.258 & 0.369&-0.072&-0.049&0.506&-0.137&0.088&-0.222& ... & ... &2.543& 2.550 & 4  \\ 
40&PDS 241& B &12.176& 0.443&-0.032&-0.048& 0.575 & -0.132 & 0.152 & -0.163 & ... & ... &1.657& ... & 4 \\
41&HD 141569& A0-A3 & 7.122&0.123&1.005& 0.083&0.052&0.031&0.140&0.995&0.068 & 0.122 &2.853 & ... & \\
42&VV Ser& B &12.342 &0.618 &0.015&0.720&0.677&-0.059&0.232&0.585&-0.155  & ... &2.603& ... & \\
43&V431 Sct&  B  &12.769 & 0.805&-0.194&0.006&0.961&-0.156&0.108&-0.411& ... & ... &2.401& 2.480 &4 \\
44&HD 174571&  B  &9.109 & 0.498&-0.153&0.328&0.605&-0.107&0.038&0.093&0.049 & ... &2.558& 2.640&   \\
45&HD 179218& A0-A3  &7.084  & 0.079&0.117&1.141&0.047&0.032&0.132&1.132&0.076 &0.269 & 2.847  & ... &  \\
46&WW Vul& A-F &10.947 & 0.264&0.171&1.002&0.151&0.113&0.219&0.972&-0.013 &0.155 &2.819& ... &  \\
47&PX Vul& A-F &12.214 &0.423  &0.025&0.347&0.133&0.290&0.068&0.320&0.119 &-0.039 &2.635& ... & 2  \\
48&PDS 581&  B  &12.037 &0.511 &0.109&0.079&0.686& -0.175 & 0.316 & -0.614 & ... & ... &2.120& 2.370& 4  \\
49&HD 190073& A0-A3 & 7.619 &0.090&0.097&1.059&0.015&0.075&0.102&1.056&0.068 &0.590 &2.680   & ... & 1  \\
50&V1685 Cyg&  B &11.506 &0.597&-0.119&-0.219&0.768&-0.171&0.122&-0.565& ... & ... &2.384& 2.400 &4  \\
51&HD 200775&  B &7.415 &0.319&-0.008&0.102&0.449&-0.130&0.132&-0.143& ... & ... &2.499& 2.580 &4  \\
52&HD 203024&  A-F &8.652 &0.121&0.138&0.906&0.006&0.115&0.140&0.905&0.067 &0.086 &2.820 & ... &  \\
53&BD +65.1637&   B  &10.151 &0.356&-0.047&0.210&0.474&-0.118&0.102&-0.022&-0.020 & ... & 2.528   & 2.610 &  \\
54&V1578 Cyg&  B  &10.124 &0.296&0.071&0.977&0.324&-0.028&0.175&0.912&-0.066 & ... &2.687    & ... &  \\
55&BH Cep&  A-F  &11.217 &0.406&0.141&0.518&0.150&0.256&0.189&0.488&-0.017 &-0.025 & 2.695   & ... & \\
56&SV Cep& A0-A3 &11.083 &0.257&0.089&0.940&0.196&0.061&0.139&0.909&0.067 &-0.025 & 2.885  & ... & \\
57&V1080 Tau& G & 11.186 &0.828 &-0.087&0.546& ... & ... & ... & ... & ... & ... &2.567& ... & 4  \\
58&CO Ori&  G  &11.452  &0.540 &0.287&0.085& ... & ... & ... & ... & ... & ... &2.596&  & 4  \\
59&V1650 Ori&  A-F  &10.726  &0.436 &0.162&0.103&0.333&0.195&0.413&-0.010 &0.048 &2.638   & ...  & \\
60&RY Ori& A-F &12.416& 0.549& 0.162&0.434&0.338 & 0.387 & 0.390 & 0.085 & -0.215 & -0.325 &2.658& ...  & \\
61&HD 36910&  A-F &10.334 &0.429&0.214&0.781&0.244&0.185&0.292&0.732&-0.097 &-0.004 &2.778& ...  & 7 \\
62&HD 53240&  B  &6.120 &-0.008&0.087&0.844&0.027&-0.035&0.096&0.839&0.006 & ... &2.671   & ...  & \\
63&HD 261387& A0-A3 &10.534  &0.093  &0.181&0.971&0.027&0.066&0.190&0.966&0.017 &0.047 & 2.874  & ...  & \\
\\
\hline
\end{tabular}
}
\label{tab:Dereddenedindices}
\end{table*}

To obtain the effective temperature ($T_{\rm eff}$), surface gravity (log g), and [Fe/H] values, different grids were used depending on the aforementioned spectral classification: 
NEMO-2004 CGM grids \citep{2004CoAst.144...43N} for A-F stars (5500K $\leq$ $T_{\rm eff}$ $\leq$ 8500K, late group), the grids of \citet{1985MNRAS.217..305M} with the log g corrections suggested by \citet{1993A&A...268..653N} 
for A0-A3 stars (8500K $<$ $T_{\rm eff}$ $\leq$ 11000K, intermediate group), and the \citet{1993A&A...268..653N} grids for the B group 
(11000K $<$ $T_{\rm eff}$ $\leq$ 35000K, early group). Table~\ref{tab:FundParI} lists the
fundamental physical parameters for the sample stars: name, absolute magnitude, effective temperature, log g, metallicity, spectral type group, and calibration used. Typical error bars for log g are about 0.25 dex for stars with 
$T_{\rm eff}$ > 20 000 K, $\approx$ 0.15 dex for stars with 11 000 $\leq$ $T_{\rm eff}$ $\leq$ 20 000 K, and $\approx$ 0.10 dex for $T_{\rm eff}$ $\leq$ 11 000 K. In the case of $T_{\rm eff}$, relative errors
in its determination are about 4$\%$ for $T_{\rm eff}$ > 20 000 K, 3$\%$ for 11 000 $\leq$ $T_{\rm eff}$ $\leq$ 20 000 K, and 2.5$\%$ for $T_{\rm eff}$ $\leq$ 11 000 K \citep{1993A&A...268..653N}. 
Typical errors in the determination of M$_{V}$ and [Fe/H] are about 0.$^{m}$3 \citep{1975AJ.....80..955C,1978AJ.....83...48C,1979AJ.....84.1858C} and 0.1 dex 
\citep{1988A&A...199..146N,1993A&A...274..391S}, respectively. \\

Although the dereddening procedure and photometric calibrations used are a well-proven method to obtain the stellar parameters for MS stars, they might not be correct for some HAEBE stars 
depending on the amount of circumstellar material. For this reason, the stellar parameters obtained must be considered just as an initial estimate and need to be checked with future spectroscopic observations. \\

\begin{table*}
\caption{Fundamental physical parameters for the sample of observed stars.} 
\centering
\scalebox{0.88}{
\begin{tabular}{cccccccc}  \hline
ID &Name& M$_{V}$ & $T_{ef}$ & log \textit{g} & [Fe/H] & Spectral & Calibration \\
number &    & (mag) & (K)      &  ~             &        & type group    & adopted \\ 
\hline
 &      &$\pm$0.3&                &                  &$\pm$0.1&   &   \\
1&VX Cas& 1.5  & 9850 $\pm$ 250  & 4.07 $\pm $0.10  & -  & A0-A3 & 3 \\ 
2&V594 Cas& -4.1  & 30500 $\pm$ 1000 & 4.05 $\pm$ 0.25  & -  & B  & 2  \\ 
3&PDS 004& 1.9   & 8030 $\pm$ 200 & 4.18 $\pm$ 0.10  & -0.10  & A  & 1  \\ 
4&XY Per& -   & -  & -  & -  & -  & - \\
5&AB Aur& 0.0   & 8780 $\pm$ 200 & 3.64 $\pm$ 0.10  & -0.65 &  A0-A3 & 3  \\
6&HD 31648&  1.5  & 7870 $\pm$ 200  & 4.02 $\pm$ 0.10  & -0.19  &  A  & 1  \\ 
9&HD 35187& 2.1   & 7490 $\pm$ 200 & 4.07 $\pm$ 0.10  & -0.29  &  A  & 1  \\ 
10&HD 287823& 1.8   & 8740 $\pm$ 200  & 4.35 $\pm$ 0.10  & -0.18  & A0-A3  & 3  \\ 
12&HD 290409& 1.7   & 9440 $\pm$ 250 & 4.27 $\pm$ 0.10  & -  &  A0-A3   & 3  \\ 
13&HD 290500& 3.6   & 7700 $\pm$ 200  & $>$4.50  & -0.46  &  A   & 1  \\ 
14&V1409 Ori& 0.8   & 7820 $\pm$ 200  & 4.19 $\pm$ 0.10  & -0.02  &  A  & 1 \\ 
15&HD 36112& 2.1   & 7770 $\pm$ 200  & 4.16 $\pm$ 0.10  & 0.02   &  A  & 1  \\
16&V1410 Ori& 1.4   & 8140 $\pm$ 200  & 4.06 $\pm$ 0.10  & -0.02  &  A  & 1  \\ 
17&HD 36408& -1.6   & 11950 $\pm$ 350  & 3.22 $\pm$ 0.15  & -  &  B  &  2  \\ 
18&V1271 Ori& -0.1   & 9310 $\pm$ 250  & 3.56 $\pm$ 0.10  & -  &  A0-A3   & 3  \\ 
19&V380 Ori&  -   & -  & -  & -  & -  & - \\
20&HD 290770& -1.1  & 11300 $\pm$ 350  & 3.41 $\pm$ 0.15  & -  & B   &  2  \\
21&BF Ori& 0.1   & 8600 $\pm$ 200  & 3.67 $\pm$ 0.10  & 0.35  &   A0-A3   &  1  \\
22&HD 37357& 1.2   & 8790 $\pm$ 200  & 4.09 $\pm$ 0.10  & -0.25  &  A0-A3  &  3  \\
23&V1247 Ori& 0.7   & 7060 $\pm$ 200 & 3.64 $\pm$ 0.10  & 0.55  &  F  &  1  \\
24&V1788 Ori& 1.4   & 8470 $\pm$ 200  & 4.16 $\pm$ 0.10  & -0.22 &  A0-A3  &   1  \\
25&RR Tau& -   & -  &  - & -  &  - & - \\
26&V350 Ori& 2.1 & 8500 $\pm$ 200 & 3.74 $\pm$ 0.10  & 0.91  &  A0-A3   &  1  \\
27&HD 37806&  -1.6  & 11500 $\pm$ 350  & 3.18 $\pm$ 0.15  & -  &  B   & 2  \\
28&HD 38120& -0.9   & 10700 $\pm$ 250  & 2.66 $\pm$ 0.10  & -  &  B   & 3  \\
30&HD 249879& -0.3   & 11350 $\pm$ 350 & 3.68 $\pm$ 0.15  & -  & B   &  2  \\
31&V791 Mon& -3.2   & 25600 $\pm$ 1000  & 4.07 $\pm$ 0.25  & -  &  B   &  2  \\ 
32&HD 250550& -3.0  & 12800 $\pm$ 350  & 2.81 $\pm$ 0.15  & -  & B   &  2  \\
33&AE Lep& 1.9   & 8050 $\pm$ 200 & 4.18 $\pm$ 0.10 & -0.84  &    A    &  1  \\
34&PDS 126& 2.5   & 8650 $\pm$ 200  & 4.50 $\pm$ 0.10  & 0.18  &  A0-A3  &  1  \\ 
35&HD 50083& -   & -  & -  & -  & -  & - \\ 
36&HD 52721& -4.7   & 33800 $\pm$ 1500  & 4.21 $\pm$ 0.25  & -  &  B   &  2  \\
37&HT Cma&0.6  & 10560 $\pm$ 250  & 3.61 $\pm$ 0.10  & -  & B & 3  \\
38&HU Cma& -1.3  & 16000 $\pm$ 500  & 4.15 $\pm$ 0.15  & -  & B &2  \\
39&HD 53367& -  & -  & -  & -  & -  & - \\
40&PDS 241& -  & -  & -  & -  & -  & - \\
41&HD 141569& 0.4  & 9220 $\pm$ 250  & 3.93 $\pm$ 0.10  & -0.64  & A0-A3 & 3  \\ 
42&VV Ser& -5.5  & 12150 $\pm$ 350 & 2.06 $\pm$ 0.15  & -  & B & 2  \\
43&V431 Sct& -  & -  & -  & -  & -  & - \\
44&HD 174571& -2.4  & 22700 $\pm$ 1000  & 4.13 $\pm$ 0.25  & -  & B & 2  \\
45&HD 179218& -0.3  & 8930 $\pm$ 200  & 3.55 $\pm$ 0.10  & -0.72  & A0-A3 & 3 \\
46&WW Vul& 1.2  & 7770 $\pm$ 200  & 3.87 $\pm$ 0.10  & 0.22  &  A & 1  \\ 
47&PX Vul& 4.5  & 6120 $\pm$ 150  & $>$4.50  & -1.18 & F  & 1  \\ 
48&PDS 581& -  & -  & -  & -  & -  & - \\
49&HD 190073& -  & -  & -  & - & -  & -  \\ 
50&V1685 Cyg& -  & -  & -  & -  & -  & - \\
51&HD 200775& -  & -  & -  & -  & -  & - \\
52&HD 203024& 1.8  & 7820 $\pm$ 200  & 4.09 $\pm$ 0.10  & -0.63  & A & 1  \\
53&BD +65.1637& -3.6  & 27200 $\pm$ 1000  & 3.99 $\pm$ 0.25  & -  & B & 2  \\
54&V1578 Cyg& -1.0 & 10830 $\pm$ 250  & 2.66 $\pm$ 0.10  & -  & B & 2 \\
55&BH Cep& 3.6  & 6750 $\pm$ 150  & 4.50 $\pm$ 0.10  & 0.36  & F & 1  \\
56&SV Cep& 1.5  & 8830 $\pm$ 200  & 4.39 $\pm$ 0.10  & -0.63  & A0-A3 & 3  \\ 
57&V1080 Tau& -  & -  & - &  - &  - & - \\
58&CO Ori& -  &  - &  - &  - & -  & - \\
59&V1650 Ori& 3.6  & 6170 $\pm$ 150 & 4.25 $\pm$ 0.10  & 0.23  & F & 1  \\
60&RY Ori& 7.1  & 5810 $\pm$ 150 & $>$4.50  & 2.72  & F & 1  \\
61&HD 36910& 2.8  & 7230 $\pm$ 200  & 4.11 $\pm$ 0.10  & 1.10  & F & 1  \\
62&HD 53240& -3.1  & 11100 $\pm$ 350  & 2.57 $\pm$ 0.15  & -  & B & 2  \\
63&HD 261387& 1.6  & 8400 $\pm$ 200  & 4.23 $\pm$ 0.10  & -0.10  & A0-A3 & 1  \\
\\
\hline
\\
\end{tabular}
}
\begin{list}{}{}
\item[]\scriptsize{\textbf{Calibrations adopted:} }
\item[]\scriptsize{1: NEMO-2004 CGM \citep{2004CoAst.144...43N,2002A&A...392..619H}. Late group, the parameters \textit{(b - y)$_{0}$} and \textit{c$_{0}$} are used here as temperature and luminosity indicators, respectively.}
\item[]\scriptsize{2: \cite{1993A&A...268..653N}. Early group, the parameters \textit{c$_{0}$} and $\beta$ are used as temperature and luminosity indicators, respectively.}
\item[]\scriptsize{3: \cite{1985MNRAS.217..305M}. Intermediate group, the parameters a$_{0}$ y r$^{*}$ together with a log\textit{g} correction \citep{1993A&A...268..653N} are used for this group.}
\end{list}
\label{tab:FundParI}
\end{table*}

\begin{figure*}
\centering
 \includegraphics[width=12cm,angle=-90]{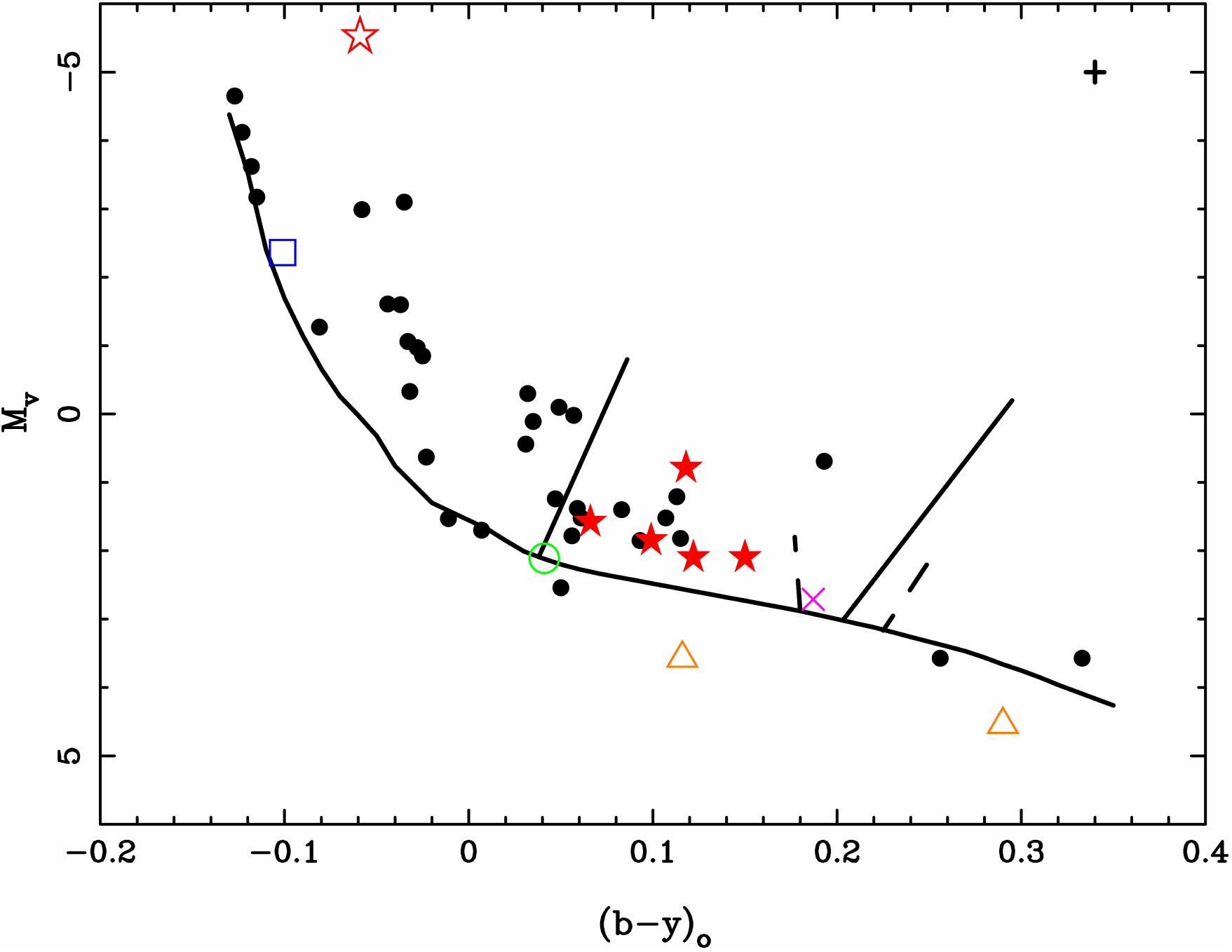}
  \caption{Colour-magnitude diagram of the sample of observed stars. Observational edges for the MS $\delta$ Scuti instability region (solid lines) are from \citet{2001A&A...366..178R} and those 
   for the $\gamma$ Dor region (dashed lines) are from \citet{2002MNRAS.333..251H}. Red solid asterisks: PMS $\delta$ Scuti-type stars (PDS 004, HD 35187, V1409 Ori, HD 36112, and HD 261387); 
   black solid circles: constant stars; green open circle: V350 Ori ($\delta$ Scuti-type Am star); blue open square: HD 174571 (probable $\beta$ Cephei-type pulsator); red open asterisk: VV Ser ($\delta$ Scuti-type star, see text for explanation);
   orange open triangles: subluminous stars (HD 290500 and PX Vul); and magenta cross: HD 36910 (Am star). Upper right: error bars.} 
         \label{fig:DiafotHR}
   \end{figure*}

Figure~\ref{fig:DiafotHR} shows a colour-magnitude diagram, where our sample of observed PMS and HAEBE stars has been placed. The observational
instability region for MS $\delta$ Scuti-type stars \citep{2001A&A...366..178R} with that of $\gamma$ Dor variables \citep{2002MNRAS.333..251H} are also shown. Zwintz (2008) 
compared the observational instability regions for pulsating PMS and classical (post- and MS) $\delta$ Scuti-type stars, making use of the sample of PMS pulsators available up to 
that date, and concluded that the hot and cool borders of both instability regions seem to coincide. Our results seems to be also in good agreement with this finding, as we show below. \\

All five PMS $\delta$ Scuti-type stars detected in this work (PDS 004 (ID 3), HD 35187 (ID 9), V1409 Ori (ID 14), HD 36112 (ID 15), and V350 Ori (ID 26)) are placed within the 
limits of the observational instability strip for the classical $\delta$ Scuti stars. We note that V350 Ori is located just on the locus of the MS and very close to the blue edge. However, 
it has been classified as an Am star in Table~\ref{tab:Dereddenedindices}. \citet{2001A&A...366..178R} showed that absolute magnitudes obtained with photometric calibrations for Am stars have 
large systematic errors, underestimating their luminosities in relation to those obtained using distances from HIPPARCOS \citep{1997A&A...323L..49P} in all cases. Therefore, V350 Ori must be 
located above the MS and slightly shifted to the left of the blue edge.\\ 

The objects HD 261387 (ID 63) and VV Ser (ID 42) are the two stars with confirmed $\delta$ Scuti-type pulsations. The former is also placed inside the classical $\delta$ Scuti 
instability region, but the latter, according to our photometric data and calibrations, is placed in the B supergiant region. This is probably due to an anomalous $\beta$ index derived from 
its variable H$_{\beta}$ line in emission \citep{1988A&A...197..151C}. However, these authors classified VV Ser as an A2e$\beta$ star by means of spectroscopic observations, which
place VV Ser within the limits of the $\delta$ Scuti region. \\

Two of the stars with detected $\beta$-Cephei-like variability, V1685 Cyg (ID 50) and HD 50083 (ID 35), are not shown in Fig. \ref{fig:DiafotHR} because 
their $\beta$ index, which is used to derive their temperatures, falls outside the limits of the $\beta$ calibration \citep{1980A&AS...40..199P}. If used, they should be placed in the B0-B2 region. 
Concerning HD 174571 (ID 44), this star could be located on the colour-magnitude diagram near the MS and in the early B stars region. 
Thereby, these three stars are located in the region, where $\beta$ Cephei-type pulsation can be excited, supporting the possibility that their membership to that group of pulsators is real. \\

Some stars that do not have any type of pulsation detected present problems concerning their photometric calibrations. The stars HD 53367 (ID 39), PDS 241 (ID 40), V431 Sct (ID 43), HD 200775 (ID 51), PDS 581 (ID 48), V1080 Tau (ID 57), and CO Ori (ID 58) could not be 
calibrated because their $\beta$ indices are outside of the range covered by the calibrations used. The first five objects are early B stars, whereas V1080 Tau and CO Ori are G stars. Moreover, PDS 241 and V380 Ori 
present anomalous $\beta$ indices very probably due to strong emission in the H$_{\beta}$ line (1.657 and 2.359, respectively), which makes it impossible to obtain any reliable $\beta$ index. \\

On the other hand, XY Per (ID 4) and HD 190073 (ID 49) are classified as high luminosity stars. There are no appropriate dereddening equations for this type of star, as it
has been already indicated by \citet{1980A&AS...40..199P}. Therefore, these stars were rejected from the calibration. The stars HD 290500 (ID 13) and PX Vul (ID 47) appear as subluminous stars due to an insufficient 
derredening in their photometric indices which is probably caused by the large amount of circumstellar dust present around these stars. Finally, HD 36910 (ID 61) is classified as an Am star located on the MS 
and close to the red edge of the $\delta$ Scuti instability strip. However, its position would be shifted up above the MS because the luminosities obtained by photometric calibrations 
for Am stars are systematically underestimated, as in the case of V350 Ori. 

\section{Summary and conclusions}

A systematic search for $\delta$ Scuti-type pulsations was carried out over a well-defined sample of northern field stars previously classified as either PMS or HAEBE stars. 
We have detected five new PMS $\delta$ Scuti-type stars (PDS 004, HD 35187, V1409 Ori, HD 36112, and V350 Ori) and three probable $\beta$ Cephei-type 
pulsators (HD 174571, V1685 Cyg, and HD 50083). In addition, three new MS $\delta$ Scuti-type pulsators (HD 202901, HD 203573, and HD 35909) and two MS stars that show 
$\gamma$ Doradus-type pulsations (HD 37594 and HD 52343) have also been detected among the comparison and check stars. Finally, two stars (HD 261387 and VV Ser) 
have also been confirmed as PMS $\delta$ Scuti-type pulsators in this work.\\

The sample of observed stars has been placed in a colour-magnitude diagram, and their stellar physical parameters were derived using Str\"{o}mgren photometry and photometric calibrations. 
All five PMS stars that have detected $\delta$ Scuti-type pulsation are located within the observational limits of the observational instability strip for the classical $\delta$ Scuti stars, 
which confirms their nature as PMS $\delta$ Scuti-type pulsators. Although photometric calibrations are a well-proven method for MS stars, it is important to note that they are not very reliable for some PMS field stars due to their 
HAEBE nature, which results from the following: emission of the H$_{\beta}$ line due to hot gas located in their accretion discs (affecting their $\beta$ index), presence of a large amount of dust in the circumstellar 
material (resulting in peculiarities in their colour (\textit{b-y}) indices), and high accretion rates that produce a luminosity excess in UV (affecting 
their \textit{c}$_{1}$ index associated to this band) in some cases. Therefore, it is important to obtain spectroscopic observations for these stars to check the physical parameters obtained. Finally, this search for $\delta$ Scuti-type pulsations among PMS stars 
is only a first step to a proper study of their inner layers. With this purpose, it is necessary to obtain new observations of long runs for each star
to obtain a more detailed frequency content that will allow to model their inner structure.

\begin{acknowledgements}
This work is supported by the Spanish Ministry of Economy and Competitiveness (MINECO) under the grant BES-2007-15372. This work has been partially funded by the projects AYA2006-06375, 
AYA2009-10394, and AYA2011-30147-C03-01 of the Spanish Ministry of Economy and Competitiveness (MINECO), cofounded with FEDER funds, and 2011 FQM 7363 of the Consejería de Economía, Innovación, Ciencia y 
Empleo (Junta de Andalucía). This work made use of the SIMBAD database, operated at CDS, Strasbourg, France. We want to thank an anonymous referee for their comments and suggestions which 
help us to improve this paper. We also thank A. Fern\'andez-Mart\'in for her suggestions and support.

\end{acknowledgements}

\end{document}